\documentclass[apj]{emulateapj}

\usepackage{psfig}
\usepackage{apjfonts}
\begin{document}

\newcommand{\be}{\begin{eqnarray}}
\newcommand{\ee}{\end{eqnarray}}
\newcommand{\Smin}{S_{\rm min}}
\newcommand{\Dmax}{D_{\rm max}}
\newcommand{\Msun}{M_{\odot}}
\newcommand{\Lsun}{L_{\odot}}
\newcommand{\Mearth}{M_{\oplus}}
\newcommand{\ngj}{n_{\rm GJ}}
\newcommand{\rhogj}{\rho_{\rm e, GJ}}
\newcommand{\Epar}{E_{\parallel}}
\newcommand{\rlc}{r_{\rm LC}}
\newcommand{\Edot}{\dot E}
\newcommand{\Pdot}{\dot P}
\newcommand{\Porb}{P_{\rm orb}}
\newcommand{\apc}{A_{\rm PC}}
\newcommand{\ExB}{{\bf E$\times$B}}
\newcommand{\EdotB}{{\bf E$\cdot$B}}
\newcommand{\OmdotB}{{\mathbf \Omega\cdot B}} 
\newcommand{\npm}{n_{\pm}}

\shortauthors{Cordes \& Shannon} 

\title{Rocking the Lighthouse:  Circumpulsar Asteroids and Radio Intermittency}
\shorttitle{Radio Intermittency from Circumpulsar Debris}

\author{
  J.~M.~Cordes \& R.~M.~Shannon
}
\affil{Astronomy Department, Cornell University, Ithaca, NY 14853}

\begin{abstract}
We propose that neutral, circumpulsar debris entering the light cylinder
can account for many time-dependent pulsar phenomena that are otherwise
difficult to explain.  Neutral material avoids propeller ejection
and injects sufficient charges --- after heating, evaporation, 
and ionization --- to alter current flows and 
pair-production and thus trigger, detune, or extinguish coherent emission.  
Relevant phenomena, with time scales
from seconds to months, include nulls, rotating radio transients (RRATs),
rapid changes in pulse profile (``mode changes''), variable subpulse
drift rates, quasi-periodic bursts from B1931+24, and torque variations.
Over the $\sim 10^7$~yr lifetime of a canonical pulsar with $10^{12}$ G 
surface magnetic field, less than 
$10^{22} B_{12}$~g ($\lesssim 10^{-6}\, M_{\oplus}$)
is needed to modulate the Goldreich-Julian current  by 100\%. 
Circumpulsar material originates from metal-rich, supernova fallback gas
that aggregates into asteroids.  Debris disks can inject sufficient material
on time scales of interest, yet be  too tenuous to form large planets
detectable in pulse timing data.  Asteroid migration results from collisions
and the  radiation-driven Yarkovsky and Poynting-Robertson effects.
For B1931+24, an asteroid in a $\sim 40$~day elliptical orbit
pollutes the magnetosphere stochastically  through collisions
with other debris.
Injection is less likely for hot, young and highly magnetized
 pulsars or millisecond pulsars
that pre-ionize any debris material well outside their small
magnetospheres.  Injection effects will therefore be most prominent
in long-period, cooler pulsars,  consistent with the distribution of relevant
objects in $P$ and $\dot P$.  A pulsar's spin history and its 
radiation-beam orientation may influence whether it displays nulling, RRATs
and other effects.
\end{abstract}


\section{Introduction}\label{sec:intro} 

Rotation-driven pulsars display myriad phenomena in their 
radio emission on time scales that range from nanoseconds to years.   
Here we consider a unifying model for phenomena that occur on time scales 
from pulse-to-pulse to month-like or longer, 
including:
(1) null pulses, where the emission evidently vanishes but returns in
pulses that maintain synchronism with the spin; 
(2) drifting subpulses, whose systematic variations in pulse phase appear 
to be related to circulation of emission regions around the magnetic pole;
(3) mode changes, where the pulse shape changes discontinuously between
two or more shapes, often accompanied by dramatic changes in flux density;
and
(4) rotating radio transients (RRATs) that correspond to bursts of
single or a few pulses spaced by very long intervals (minutes to hours) 
\cite[][]{m+06}.  A few objects show  
giant pulses that can be $\gtrsim 10^3$ larger in amplitude than
the average pulse \cite[e.g.][]{cstt96, hwek03, jr03, cbh+04, c04}. 

\begin{figure}[!ht]
\begin{center} 
\includegraphics[scale=0.45,angle=0]{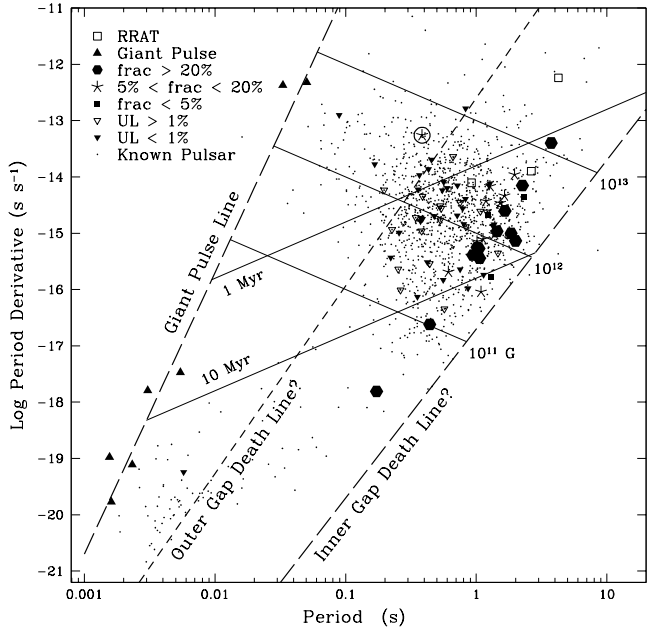}
\caption{\label{fig:ppdot} The period-period-derivative diagram for pulsars. 
Rotating radio transients (RRATs),
pulsars that show giant pulses, and nulling pulsars that display different
nulling fractions are distinctly labelled (see legend).  
Points labelled ``UL'' indicate pulsars for which upper limits have 
been placed on the nulling fraction.  
Solid lines of constant magnetic field and lines of constant
characteristic spindown age, $P / 2\Pdot$ are shown.
The most leftward dashed line
denotes constant magnetic field strength at the light cylinder 
(with a field strength equal to that of  the Crab pulsar) that
seems to be a trademark of objects showing giant pulses 
\cite[e.g.][]{cstt96}.
The inner-gap ``death line''  is drawn as the rightward dashed line
 at an arbitrary location that is
consistent with the diminishment of the radio pulsar density in the diagram.  
The middle dashed line is an outer-gap death line from
\cite{t+06}.
The pulsar B0656+14 is circled because it is an object with high
surface temperature that shows low-level nulling.} 
\end{center}
\end{figure}

Coherent emission wanes on time scales $\sim 10^{7}$ yr for canonical
pulsars --- those with surface magnetic fields $\sim 10^{12}$ G --- and 
on much longer times for millisecond pulsars (MSPs; $10^8 - 10^9$ G).  
We assume, of course, that the spindown time scale, $\tau_s = P / 2\dot P$, 
is a reasonable, but by no means perfect, proxy for chronological age.  
The distribution of pulsars in the period-period derivative 
($P$-$\dot P$) diagram shown in Figure~\ref{fig:ppdot} 
displays a dearth of pulsars with 
$\dot P < 10^{-16.7}\,{\rm s\,s^{-1}} P^3$
in the bottom right corner 
that reflects these time scales.  
However, it is
not clear whether pulsars shut off their radio emission according to
a threshold effect, such as the inability to sustain
electron-positron cascades whose flows induce coherent emission, or
that the rotational energy loss rate, $\dot E = I \Omega\dot\Omega$, simply
cannot provide sufficient power to drive the radio beam luminosity 
\cite[]{acc02}.   
Empirically, the latter seems favored because the density of pulsars 
in the $P-\dot P$ diagram rolls off smoothly rather than 
sharply at a ``cliff''.  However,  pair cascades may terminate differently in 
pulsars with the same $P$ and $\dot P$ if surface magnetic fields show 
object-dependent complexity, thus smearing out any cliff effect.        

In this paper we develop a mechanism that pertains to the
four phenomena described above but is particularly
 focused on nulling pulsars and RRATs. 
Both  show episodes of radio
quiesence that appear to be totally devoid of any radio emission 
to levels $\lesssim 10^{-2}$ of their `on' states.    
Off states can last for long time periods (from minutes to days), although
some nulling pulsars are
off only for a few pulse periods.  
Both classes of objects are largely found on the right-hand periphery of
the population of canonical pulsars in the $P-\dot P$ diagram 
(Figure~\ref{fig:ppdot}).   As pointed
out by many \cite[e.g.][]{b92}, nulling may signify a faltering
emission process that signifies incipient complete shut off of radio emission.   
Instead, we 
suggest that nulling and RRATs are both caused by pollution
of NS magnetospheres by low-level accretion 
from a circumpulsar asteroid disk.
This idea is motivated by the fact that one nulling pulsar, B1931+24, has 
state durations of days that are cyclical with a
$\sim 40$-day quasi-period \cite[][]{k+06}. 
In addition, early work by \cite{c85} 
explored the role played by interstellar grains in the current flows
of active pulsars and their manifestation on the spindown rates of pulsars.
Our model differs by considering material that originates in a reservoir
of orbiting debris, most likely supernova fallback material that has
formed a low-mass asteroid belt.   Orbits around neutron stars provide
a plethora of time scales that can be imprinted 
on electrodynamic processes by material injected into the magnetosphere.
\cite{md81} considered plasma disks around pulsars to address
issues of electrodynamics in pulsar magnetospheres.   Here too, our
model is different in considering a largely neutral disk comprising
macroscopic objects.

Cheng's (1985) grain inflow model was developed to explain an apparent
correlation between $P \dot P$ and space velocity that was identified
in the 1980s.   While that correlation is now thought to be insignificant
for canonical pulsars\footnote{Note that space velocities are much less
for pulsars that have undergone accretion-induced spinup and consequent
diminishment of the dipole field component.  However, there does not
appear to be a correlation between $B_{\rm dipole}^2\propto P\dot P$ 
and space velocity among pulsars that have not accreted much material.},
extrinsic charges from inflowing grains
also solved the issue of how a NS avoids being charged up
by self-generated current flows and it also obviated the need for
self-sustained pair production.   However,   the grain model appears to
work only for cool ($\sim 10^5$ K), long-period pulsars because
grains are evaporated by thermal radiation at distances of
$10^9 - 10^{10}$~cm, comparable to the light-cylinder radii of
long-period pulsars, but much larger than those of short-period 
and, presumbably, younger and hotter pulsars.  For our
purposes, the longer-period and (presumably) cooler NS are just the sample
we wish to address.  In many respects, Cheng's analysis also applies to
grains delivered from asteroids that are evaporated and ionized
inside the magnetosphere.  The differences are that
(a) orbital delivery is expected to be episodic, as opposed to the
steady flow of interstellar grains; (b) asteroids and rocks
will penetrate further into the magnetosphere than the 
$\sim 0.1\,\mu m$ grains considered by Cheng; and (c) circumpulsar disks
probably vary widely from pulsar to pulsar, providing a natural explanation
for why pulsars with very similar spin parameters ($P$, $\dot P$) are
dramatically different with respect to the occurrence of nulls, etc.    


In the next sections, we describe relevant phenomena
and key elements of the model for producing 
nulls, RRATs and other bursts, including 
(a) formation of the circumpulsar disk;
(b) delivery of asteroids from the disk to the magnetosphere;
(c) tidal disruption,  evaporation, and ionization;
and
(d) effects on gap accelerators and radio emission. 
We also discuss implications for spin variations of NS from the
torque fluctuations that are expected and we suggest possible 
tests for the model. 

\section{Salient Properties of Nulls and RRATs}\label{sec:salient}

The most striking feature of both nulls and RRATs is the
rapid transition between states, either on to off or vice
versa.  These transitions appear to occur in less than one spin period in many cases   
\cite[e.g.][]{d+86} and are thus too fast to be caused by
propagation effects in the interstellar medium (ISM).   
The nulling phenomenon includes
pulsars that are mostly on and occasionally show nulls with durations
of a few spin periods and others where the pulsar is mostly 
in the off state for periods of up to days.        
In a few well studied objects, a slow roll-off 
of the pulsed flux precedes a null.  In one case
 (B1944+17; \cite{d+86}) the slow decay is over a few pulse periods and
in another, a few tens of periods (J1752+2359; \cite{l+04}).  
Nulling objects occur throughout the grouping of canonical pulsars
in the $P-\dot P$ diagram.   However, those with null fractions
$> 20$\% are strikingly more common
near the right-hand side of the grouping.  Notable outliers exist, such as B0656+14, as labelled in Figure~\ref{fig:ppdot}.   Data used
to form Figure~1 are given in Table~\ref{tab:nulldata} and are from
multiple sources in the literature and also from
the ATNF/Jodrell pulsar catalog 
(\url{http://www.atnf.csiro.au/research/pulsar/psrcat/}).

Some objects with nearly the
same period and period derivative null while others do not.
This suggests that $P$ and $\dot P$ are not the sole determinants of
a pulsar's instantaneous radio luminosity.   
The surface magnetic field strength is
commonly estimated to be $B_{12} = (P \dot P_{-15})^{1/2}$ (with
$\dot P = 10^{-15}\dot P_{-15}$) and numerous work has associated
either the pseudo-luminosity $S D^2$, where $S$ is the flux density
averaged over period and $D$ is the distance in kpc, or the true
beam luminosity \cite[e.g.][]{acc02} with a power-law model
$P^x \dot P^y$, with exponents $x,y$ determined from some fitting process.
The presence and absence of nulling objects with the same spin parameters
suggests that other variables are involved.  Intrinsic models might
include instabilities in the current flow that are regulated
by the thermal time constant of the NS crust \cite[][]{cr80, j82} or
incomplete filling of the open field line region with emission
regions associated with drifting subpulses \cite[e.g.][]{dr99}. 

\begin{figure}[!ht]
\begin{center}
\includegraphics[scale=0.45, angle=0]{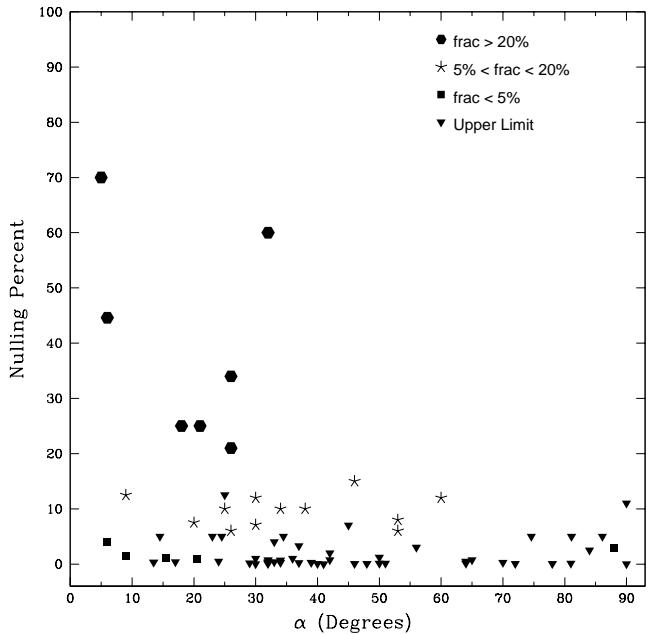}
\caption{\label{fig:null_angle}Pulsar nulling fraction as a function 
of inclination angle $\alpha$ between the magnetic and spin axes.  
The model presented in this paper suggests 
that orthogonally rotating ($\alpha \approx 90^{\circ}$) would be 
less likely to null, as the beam would severely disrupt asteroid transport 
into the magnetosphere, which is consistent with these observations.}
\end{center}
\end{figure}

Figure \ref{fig:null_angle} shows the nulling fraction plotted against
the angle between the spin axis and the magnetic 
axis $\alpha$ for the known nulling 
pulsars. Spin-magnetic axis angles 
were compiled from 
the analyses of  Rankin (1990, 1993), \cite{ew01}, and 
\cite{mr02}.  With the exception of B2315+21, 
which has a nulling fraction of $3.0 \pm 0.5 \%$\cite[][]{b92} , all the nulling pulsars 
have a spin axis-magnetic axis offset of at most $60^{\rm o}$.       

RRATs are perhaps an extreme form of nulling.   Pulse event rates range
from 0.3 to 20 hr$^{-1}$ in the eleven objects reported by 
\cite{m+06}.   A twelfth object \cite[][]{c+06} 
has a much higher rate ($\sim 200$ hr$^{-1}$) was also discovered
as an aperiodic source but later found to be periodic.
  
RRAT pulses typically occur
singly, in contrast to nulling pulsars in their on states.  
Though RRATs are biased to long periods, as with nulls there is a short-period
example.  The RRAT Galactic population is probably very large in spite of
the small, known sample because their low duty cycles imply a small
probability of detection and because
the Parkes multibeam survey covered only a narrow range of Galactic
latitude.   Non-pulsed X-ray emission has been detected from
the RRAT J$1819-1458$ \cite[][]{r+06}, which has the largest apparent magnetic field
($5\times 10^{13}$ G), and has the lowest spindown age (117 kyr) of
the 11 RRATs.

\section{Basic Picture}

The essential feature of our model  is a disk of
asteroids that is sustained in largely neutral form for tens of millions 
of years from which asteroids or their fragments are injected into the 
magnetosphere of a NS.   To account for observed nulls and RRATs, injection
must be episodic but, over long time periods, statistically homogeneous.
Thus, our model differs substantially from that of \cite{c85}, who
considered steady injection of grains from the general ISM, 
requiring their survival through the bow shocks that we know are associated 
with essentially all pulsars, owing to their large space velocities,
$V_{\rm psr} \sim 10^2 - 10^3$~km~s$^{-1}$. Moreover, his model 
implies that pulsar radio emission should always be steady (or always
quenched) rather than intermittent.    Cheng's model may be relevant
for older, cooler neutron stars that have no debris disks around them.
Our model also differs substantially from that of \cite{cp81}, who
considered interstellar objects that would reach the NS surface in
producing gamma-ray bursts from Galactic NS.

A supernova fallback disk (see below) is
metal rich and can be sufficiently low mass ($10^{-6}\Msun$) or compact
(inside the tidal disruption radius) to
prevent formation of massive planets that would be detectable individually in
pulsar timing measurements.   However, rocks can form that will resist
tidal shredding until they are well inside the magnetospheres of
long-period pulsars and will stay neutral until they are melted and ionized
by radiation from the NS surface or magnetosphere.  Delivery of rocks to
the magnetosphere can occur from collisions and orbital perturbations
by protoplanets if the disk extends beyond the tidal disruption radius.
Alternatively, rock migration from a compact disk will occur from the
radiation driven Yarkovsky effect.

The situation we identify is essentially the complement of that
needed to account for planet formation around ordinary stars.
Planet formation requires that planetesimals exist outside the
gravitational tidal radius $r_{\rm tg} \sim 10^{11}$~cm (see below)
so that runaway accretion followed by oligarchic growth and mergers
of oligarchs can form planetary core masses of Earth mass or larger
\cite[e.g.][]{r03a, g+04}. The planet building process takes only
$\sim 10^5$~yr in the inner solar system but cleanup of debris can
take several hundred million years \cite[][]{g+04}.    For our case,
the disk need not be stable; in fact it can disperse 
 on a time scale $\sim 10-100$ Myr, the apparent radio-emitting
lifetime of a canonical pulsar.

\begin{figure}[!ht]
\begin{center}
\includegraphics[scale=0.45, angle=0]{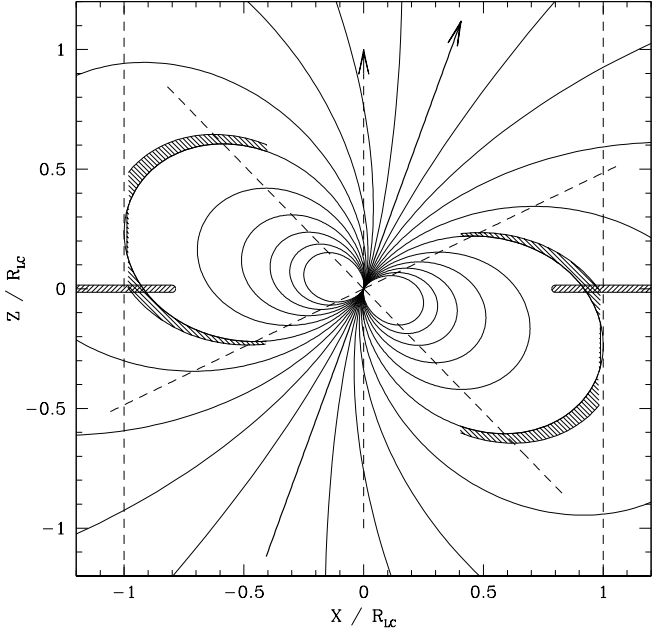}
\caption{\label{fig:cartoon}
The magnetosphere of a pulsar showing the locations of 
outer-gap acceleration regions where $\Epar\ne0$.  The asteroid belt is shown (shaded) in the $z=0$ plane.  The other shaded regions show outer gaps whose extents along and transverse to 
the last closed  field line within the light cylinder (LC) 
are arbitrary but consistent with discussion 
in the literature \cite[e.g.][]{t+06}.   These gaps may extend all the way from the LC (denoted as vertical dashed lines near
the horizontal boundaries of the diagram)  to the NS
surface, and the inner gap (too small to be shown).
The spin axis is the arrowed (vertical) dashed line and the magnetic axis is
shown by a arrowed (slanted) solid line.   
  Length scales 
are in units of the LC radius.
}
\end{center}
\end{figure}

Once a disk becomes neutral, asteroid migration 
will follow from many of the processes that have been identified in the 
solar system.  These include 
formation and growth of planetesimals through collisions, 
destructive collisions, 
tidal circularization and spin alignment, 
shepherding by larger bodies, 
injection into resonances, 
heating  from photon and particle fluxes and subsequent anisotropic
re-radiation, and heating from induction and tidal flexure.
Once injected into the magnetosphere of a pulsar, asteroids will evaporate
and, when ionized, provide charges that are subject to strong electromagnetic
forces.   Such forces are most prominent in ``gaps'' where there is
a potential drop along magnetic field lines that will accelerate
charges inward and outward.  Even though the structure of such gaps has been a
subject of analysis for more than three decades, 
a definitive picture has yet to emerge.
However, the prominence of pulsars at both radio frequencies and in high-energy
bands attests to the fact that such gaps exist.  In Figure~\ref{fig:cartoon}
we show a schematic magnetosphere for an object with a spin-magnetic-moment
angle $\alpha =20^{\circ}$.  The gaps shown are schematic representations
that are consistent with those in the literature but may underestimate 
their true extent if gaps extend all the way to the inner gap.   
It is plausible for the geometry shown that asteroidal material 
that penetrates the LC will be subject to gap potentials. 

Injected charges are accelerated in gap electric fields and can 
activate a quiescent region, such as an outer-gap,
 as a source of gamma-rays that pair produce,
or they can temporarily shut off the pair production of an active gap
by reducing the accelerating field.   A pair avalanche from an outer-gap
(as opposed to a sustained cascade) can influence other regions in the
magnetosphere, as suggested by  \cite{w03} and \cite{zgd06} for self-driven
processes within the magnetosphere rather than by extrinsic causes,
as we argue here.

In order for this picture to be viable, it must satisfy three constraints:
(1) the disk must be sufficiently low-mass that it not produce planets
that would be detectable in pulse timing measurements or by other means;
(2) the rate of mass inflow from the disk into the LC must be large
enough to influence magnetospheric current flows on time scales of
interest; and
(3) the disk must survive long enough (e.g. $> 10$~Myr) to influence
the electrodynamics of old pulsars.
Next we set limits on the properties of the disk with these constraints in mind.

\subsection{Observational Constraints on Disk Masses}

Infrared observations with {\it Spitzer} and previous instruments 
\cite[e.g.][and references therein]{bryden06}
have failed to detect disk emission around the MSP B1257+12, which
harbors at least three planets \cite[][]{kw03}.  Nonetheless, the 
observations still allow a debris disk 
around the MSP consisting of (e.g.) 100-km asteroids
with as much as $10^{-2}\Mearth$ total mass.  Related theoretical
work \cite[][]{mh01, bryden06} imply that asteroids larger than
$\sim 1$ km can exist for longer than 1 Gyr around the MSP.   

\cite{p93} placed limits on asteroidal masses around the Vela pulsar
from an upper limit on the unpulsed flux $S_u$ that would be 
produced by reflections 
of the pulsed radio flux $S_p$ from an asteroid belt.
His limit for a power-law size distribution
of slope $-3.5$ extending from  $R_{a2} = 100$~km  down to 
$R_{a1}$ equal to the 20 cm radio wavelength is
\be
M_a \lesssim 53\,M_{\oplus}	 
	\frac{r_{\rm AU}^2}{A_{0.1}} 
	\left(\frac{S_u/S_p}{2\times10^{-4}}\right)
	\left( \frac{\rho}{3\,g\,cm^{-3}}\right)
	\left(\frac{\sqrt{R_{a1} R_{a2}}}{141\,m} \right).
\ee    
For compact disks with $r_{\rm AU} \sim 10^{-2}$, this result becomes
modestly constraining as $M_a \lesssim 10^{-2.3}\, M_{\oplus}$ but
does not limit the mass in rocks that are smaller than
the 20 cm wavelength.

Pulsar timing measurements are sensitive to orbital perturbations
from objects of about the Moon's mass \cite[][]{kw03}.  The rms sinusoidal
timing residual caused by a single object of mass $m_a$ in 
a circular orbit with period $\Porb$ is 
\be
\delta t \approx \frac{a\sin i\, m_a}{\sqrt{2} c M_*} 
	\approx 0.85 \,{\rm ms}\, 
		\sin i\, \Porb^{2/3}
		\left ( \frac{m_a}{\Mearth} \right)
		\left ( \frac{1.4 \,\Msun}{\Msun} \right).
\ee
Limits on single planet masses depend on data set lengths but are typically
\cite[][]{tpc93} $\sim 1\, M_{\oplus}$ for $P_{\rm orb} \sim$~1~yr.
An ensemble of $N_a$ asteroids in a belt with total mass
$M_a$ with rms semi-major axis $a_{\rm rms}$ yields  an rms 
timing residual
\be
\delta t_{\rm rms} = 0.76\,{\rm ms}\,\,\frac{\zeta a_{\rm rms}\sin i}{\sqrt{N_a}}
	\left (\frac{M_a}{M_{\oplus}} \right ),
\ee
where $a_{\rm rms}$ is in AU and $\zeta\approx 2$ is the ratio of rms to mean 
asteroidal mass.\footnote{For a pulsar timing data set of
length $T$, only orbital periods $\lesssim 2\pi T$ should be included
in the calculation of $\delta t_{\rm rms}$ because a low-order polynomial
that is fitted and removed filters out the contributions
of longer-period objects.} 

Timing residuals for long-period pulsars are typically in the range
of 0.1 to 1 ms (rms) and show a variety of characteristic time scales as well
as displaying non-stationary behavior (i.e. rms residual that depends on
data span length).   The limit on the total mass in an asteriod belt
is therefore 
\be
M_a / \Mearth \lesssim (0.1 - 1)\times 
	\frac{\sqrt{N_a}}{a_{\rm rms}\sin\, i}. 
\ee
Thus disks  of $\gg 1\,\Mearth$ aggregate mass 
in $N_a \gg 1$ asteroids 
 can easily reside around many pulsars, particularly if
$a_{\rm rms} \ll 1$~AU, as we consider here.    In addition, such
disks can contribute significantly to the timing noise while not 
yielding detection of any individual asteroid.
To detect at the $\ell\times\sigma$ level the sinusoidal variation 
in a Fourier analysis of
$n_d = n_{d,3}10^3$ data points with white noise measurement errors of
$\sigma_{w,1}\,{\rm ms}$, an asteroid would have to satisfy
$a_{\rm AU}\sin i\, (m_a / \Mearth)\gtrsim0.04\ell\sigma_{w,1} n_{d,3}^{-1/2}$.
In the case where timing residuals are dominated by
asteroidal noise from $N_a$ asteroids, to be detectable 
an asteroid must have a mass that is a multiple
of the mean asteroidal mass, $\langle m_a \rangle M_a/ N_a$, of
\be
m_{\rm a, max} / \langle m_a \rangle \gtrsim 
	\ell\zeta (a_{\rm rms} / a_{\rm max}) \sqrt{N_a / n_d},
\ee
which is large for a large number of asteroids and reasonable values
of other parameters (e.g. $\zeta =2$, $a_{\rm rms}/a_{\rm max} = 1/2$,
and $n_d \approx 10^3$).  We conclude that canonical pulsars can
harbor asteroid disks with a sizable fraction of an Earth mass in the disk
and show timing noise consistent with that measured.  As shown below,
very low mass asteroid disks are sufficient to disrupt the magnetospheres
of these same pulsars.

\subsection{Magnetospheric Currents and Disk Lifetime \label{sec:obscons}}

To see if a disk can  provide mass at interesting rates after 10 Myr, 
we consider the current and implied mass rate needed to alter the electric 
field in a magnetosphere.  
In order for the acceleration region to exist, it is necessary for the magnetosphere to have  low density.   

The fiducial  charge density in such objects
is of order the Goldreich-Julian density \cite[][]{gj69},
$\rho_e = -\Omega\cdot B / 2\pi c$, corresponding  to a number
density 
\be
\ngj(r) =  \frac{\vert \Omega \cdot B(r)\vert}{2\pi Z e c} \approx
	 10^{10.8} \, {\rm cm^{-3} \,}
	\left (\frac{\vert \cos\psi\vert B_{12}} {ZP} \right ) 
	\left (\frac{r}{R_*}\right )^{-3}. 
\ee
Here $Z$ is the charge per particle, 
$\psi$ is the local angle between the spin axis and dipolar field,
and $R_*$ is the NS radius.  

The current from the magnetic polar cap (defined
as the footprint of dipolar field lines  that do not close
within the LC) with area $\apc$ is (in charges~s$^{-1}$)
\be
\dot N_{\rm GJ} = c\ngj \apc = \frac{2\pi^2 B R^3}{ZceP^2} =
	10^{30.14}\,{\rm s^{-1}} Z^{-1} B_{12} R_6^3 P^{-2}.	  
\ee      
For $\mu_p$ proton masses per charge, the corresponding mass rate is 
\be
\dot M_{\rm GJ} = 10^{6.3}\, {\rm g\,s^{-1}}\,\mu_p B_{12} R_6^3 P^{-2},
\ee
only a fraction $10^{-11.6}$ of the Eddington mass 
accretion rate for a NS.  
Note that the entire magnetosphere has a minimum mass
$M_{\rm mag} \approx 
        4\pi B_{\rm surface} R_*^3 \mu_p m_p \ln (r_{\rm LC} / R_*)/ ecP 
        \approx 10^{7}\, {\rm g}\, B_{12} \mu_p P^{-1}$.

For a pulsar born with a 10 ms spin period
that spins down to 10 s in $10^7$ yr, 
the total mass that would have to be accreted
to account for the GJ current is, if we assume constant magnetic field, 
\be
M_{\rm GJ} = 10^{21.9} \,{\rm g}\, \mu_p B_{12} R_6^3 \tau_7
	\left(\frac{P_0}{\rm 10\,ms}\frac{P_{\tau}}{10\,s} \right)^{-1},
\ee 
which is equivalent to the mass in a $\sim 100$ km sized asteroid.
\cite[][]{md81} made a similar point in their discussion of much more
massive plasma disks around pulsars. 
A pulsar cannot accrete during approximately the 
first $10^5$ yr of its life because tidal shredding and
thermal radiation will ionize
material before it can enter the magnetosphere.  However, a pulsar
spends most of its lifetime at long periods, so our estimate is
relevant.

Consider a disk that contains a total mass $M_a$ 
in $N_a$ asteroids within  minimum and maximum radii, $r_{1}, r_2 \gg r_1$, 
and thickness $2H \equiv \epsilon_h r_{\rm 2}$.
Inward migration implies a lifetime (for fixed $N_a$) 
\be
\tau_d = \frac{M_a}{\dot M_a} = \frac{N_a}{\dot N_a}.
\ee
The flux of asteroids through a cylinder centered on the NS with fiducial
radius $r$  and surface area
$A_c = 2\pi \epsilon_h  r_{\rm 2}  r$ is
$\dot N_a  = n_a A_c \vert \dot r \vert 
\approx 2 N_{\rm a} r \vert \dot r \vert  r_2^{-2}$,
implying a disk lifetime that  must be larger than the radio lifetime 
$\tau_{\rm psr}$. This yields the constraint
\be 
\frac{2r\vert \dot r\vert}{r_2^2} < \tau_{\rm psr}^{-1}.
\ee

In order to provide at least a  fraction $\epsilon_{\rm GJ}$ 
of the Goldreich-Julian mass rate, we require 
$\dot M_a \ge \epsilon_{\rm GJ} \dot M_{\rm GJ}$,
where $\dot M_a = \dot N_a m_a$ and $m_a$ is a typical asteroid mass, or
\be
\frac{2r\vert \dot r\vert }{r_2^2} > 
\frac{\epsilon_{\rm GJ} \dot M_{\rm GJ}}{M_a}.
\ee
Evaluating the two constraints assuming, while taking $\tau_{\rm psr} = 100$~Myr), we find
\be
10^{-21.5}\,{\rm s^{-1}}\, 
	\frac 	{
		  \epsilon_{\rm GJ} \mu_p B_{12} R_{*,6}^3
		}
		{ 
		   P^{2} \left(M_a/ \Mearth\right) 
		}
	<
	\frac{2r\vert \dot r\vert }{r_2^2} 
	< 10^{-15.5}\,{\rm s^{-1}}\, \tau_{\rm psr, 100}^{-1},
\ee
where we have assumed
uniform mass density $\rho$ in an asteroid.  
It is obvious that a disk can easily satisfy these constraints 
and not be detectable in pulse timing data. 
The implied migration rate $\dot r$ 
into a cylinder with fiducial  radius $r = 10^{10}r_{10}$~cm 
from a disk of outer radius $r_2 = 10^{-2}r_{2,2}$~AU must satisfy
\be
&&10^{-9.1} \,{\rm AU\,\,Myr^{-1}}\, 
 \epsilon_{\rm GJ} \mu_p B_{12} R_{*,6}^3 P^{-2} 
        \left(\frac{M_a}{\Mearth}\right)^{-1}
	r_{2,2}^2 r_{10}^{-1}
< \vert \dot r \vert \nonumber \\ 
&&\quad\quad < 10^{-3.1}\,{\rm AU\,\,Myr^{-1}}\, r_{2,2}^2 r_{10}^{-1}
		\tau_{\rm psr, 100}^{-1}.
\ee 
Influx rates much larger than these occur in the solar system due to a
variety of effects that we consider below. 

\section{Circumpulsar Disks}\label{sec:disks}


A dust disk around the anomalous X-ray pulsar 4U 0142+61,
a magnetar, has recently been detected
with Spitzer Space Telescope and is estimated to have mass 
$\sim 10\,\Mearth$ and a lifetime of at least 1 Myr \cite[][]{wck06}.
While other direct evidence for circumpulsar disks is lacking, 
there are several  
lines of argument that support their 
existence -- perhaps transitory --- around at least some NS.   
The strongest is the presence of
planets around at least two NS, one an MSP in the Galactic disk
(B1257+12) and the other in a globular cluster \cite[B1620-26,][]{si03}.  
For these cases, disks probably
arose from ablative disruption of  a companion star
and are not relevant to the model proposed here,
which concerns canonical pulsars that are now isolated.   
\cite{md81} explored how the electrodynamics of pulsars are altered 
by fossil disks and could alleviate the need for the then ``extreme''
magnetic field strengths of $10^{12}$ G as well as account for some of 
the observational properties of pulsars, including nulling.  
The disks proposed by \cite{md81} are substantially more massive
(up to $10^{-4}\Msun \approx 33\Mearth$) than those considered here. 


Fallback disks are plausible features of supernova explosions because
of the wide range of energy and angular momentum carried by material that is
initially outward-moving.
Simulations indicate that 
the fallback disk around a NS typically does not provide the conditions
needed for planet building, primarily because it is born compact
and does not spread much in radius when it becomes neutral and
essentially inviscid \cite[][]{mph01b}. Rocks with large tensile strengths
can grow from the metal-rich gas and survive well inside the Roche radius, 
$r_{\rm tg}$ (see below). 

\cite{mph01b} and \cite{lc02} consider  
metal-rich, supernova fallback disks of mass $10^{-6} - 10^{-3}\Msun$.
After a phase of relatively strong accretion onto the NS that mimics that in binary systems
and produces X-ray emission, the disk becomes neutral
on a time scale $\sim 10^3 - 10^4$ yr.    
The disks are born very compact ($\sim 10^8 - 10^9$ cm) and spread
to $10^9 - 10^{10}$ cm by the time they become neutral 
\cite[][]{mph01b}.
\cite{ea05} argue that disks are not disrupted by pulsar radiation 
pressure and may extend into the light cylinder. 
\cite{bp04} discuss modification of the spindown law by an accretion disk
and fit models to the Crab and Vela pulsars' braking indices and 
spindown ages to constrain disk parameters.
\cite{jl05}  consider how accretion from fallback disks with
masses of $10^{-6}- 10^{-2}\, \Msun$ will significantly
spin down NS during their first $10^4$ yr, thus providing
a mechanism for ``injection'' of pulsars into the Galactic populatation
over a range of periods that extends to $> 0.1$ s \cite[e.g.][]{vn81}. 

Growth of planetesimals in the disk 
depends on whether the material is interior or exterior to the tidal
disruption radius, which is, for a self-gravitating object with
mass density $\rho$ orbiting a 1.4\,$\Msun$ NS,
\be
r_{\rm tg} \approx \left( \frac{3 M_{\rm *}}{2\pi\rho}\right)^{1/3}
	\sim10^{11}\rho^{-1/3} \, {\rm cm}.
\ee
The tidal radius is well outside the light cylinder radius, 
\be
\rlc = c/\Omega = 10^{9.7}P \, {\rm cm},   
\ee
for all known radio pulsars but is comparable to the light cylinders 
of magnetars with $P \sim 10$~s. 
The spin rates of solar-system asteroids indicate that they are mostly
rubble piles with negligible tensile strength \cite[][]{Richardson2002} but asteroids
around pulsars may be stronger macroscopic objects if they are metal
rich and fractionated and thus can remain intact 
well inside the gravitational tidal disruption radius.
A meter-sized rock, $R_a = 10^2 R_2$~cm, with tensile strength 
${\cal T} = 10^{8} {\cal T}_8$~dyn~cm$^{-2}$, is tidally disrupted at   
\be
r_{\rm tm}\approx\left(\frac{G M_{\rm *}\rho R_a^2}{{\cal T}}\right)^{1/3}
	\approx  10^{7.4}\,{\rm cm} 
\left (\rho R_2^2{\cal T}_8^{-1}\right)^{1/3},   
\ee
which is about 27 times the NS radius, $R_{\rm *} \approx 10$~km, 
similar to altitudes where coherent
radio emission is produced \cite[][]{drh04}.  Thus, should     
any large $\sim 10$~km-sized planetesimals exist in the disk, they  
will tidally shred near the LCs of 1-s pulsars, while meter-sized rocks 
can penetrate deeply into the magnetosphere until they evaporate.

\section{Heating of Asteroids}

Heating of asteroids will ultimately evaporate grains and allow them 
to be ionized efficiently by radiation from the NS and magnetosphere.
\cite{c85} 
concluded that $\sim 0.1\,\mu m$ iron and silicate grains reach distances
$\sim 1- 3 \times 10^9T_5$cm.   Depending on their sizes and inward
velocities, meter (or larger) sized rocks will reach similar  radii
before they are evaporated.

The equilibrium temperature of an asteriod
determines whether or not it can enter the magnetosphere intact and
in a largely neutral state.
The temperature of a low-albedo asteroid in thermal 
equilibrium with surface radiation from the NS is 
\be
T_{a,*} = T_{\rm *} \left(\frac{R_{\rm *}}{2r}\right)^{1/2} 
	\approx 707\,K\,T_5 r_{10}^{-1/2},
\label{eq:Ta}
\ee
where $T_{\rm *}$ is the effective radiation temperature that ranges from $\sim 10^5$ to a  few times $10^6$ K.   Cooling
mechanisms are not well identified and recent work \cite[][]{shm02}
yields an upper limit of $10^6$ K for the young NS 3C58 while 
some older pulsars show higher temperatures.   
Moreover, NS mass determinations now
show a broader range, extending from 1.3 $\Msun$ to as high as 
2.1 $\Msun$ \cite[][]{n+05}, which signifies that a broad range of surface
temperatures should be expected for objects of the same age
\cite[][]{yp04}  because cooling is much more efficient for high-mass
NS.

Heating from magnetospheric
radiation is mostly by beamed X-rays with luminosity
$L_X \approx 10^{-3}\dot E$ \cite[][]{becker97},  
\be
T_{a,m} 
	& \approx &
	612 \, K\, \left (g_b I_{45} \dot P_{-15} P^{-3} \right )^{1/4}
	\left( \frac{L_x/\dot E}{10^{-3}}\right)^{1/4} r_{10}^{-1/2},
\ee
where the beaming factor $g_b \approx (8/\pi\theta_b)^{1/4}> 1$ 
if the beam of size $\theta_b$ (FWHM) illuminates the asteriod and 
zero if it does not.

An additional source of heating is ohmic dissipation (induction heating),
discussed in the Appendix.  The dominant effect is from current flow
between the asteroid and magnetosphere. 
The regime is different from the induction heating of asteriods in
the solar nebula during the Sun's T-Tauri phase 
\cite[e.g.][]{s+70, h89}, where
the ambient plasma density in a much stronger (than present)
solar wind provides
a current circuit through the asteroid that causes ohmic heating.  In the
pulsar case, the asteriod is in a near vacuum, at least near the light
cylinder, so charge carriers are available primarily from the
asteriod itself.
In moving closer to the NS, evaporation from radiative heating 
becomes significant (see below), providing a large number of charges.  
During this phase, 
runaway ohmic heating will occur
(R.~V.~E. Lovelace, private communication)  
even if,  as in the solar nebula case, ohmic heating 
is limited by diagmagnetism from current-induced magnetic fields 
\cite[][]{s+70}.  At 2000~K, which an asteroid reaches at
$r\sim 10^{9}$~cm around a $10^5$~K and $10^{12}$~G NS, 
evaporation begins to provide sufficient
charges for significant ohmic heating, leading to  runaway heating that
causes the asteroid to explode.   Lower magnetic fields allow the asteroid to
reach slightly smaller radii before exploding.   Perhaps the only way
to get asteroids to significantly smaller radii is for them to
intrude along the spin axis of pulsars with nearly aligned 
magnetic moments.    The minimum radius reached is perhaps irrelevant as
long as charged particles are injected inside the magnetosphere  where
they are electromagnetically captured.    

Finally, tidal heating of an object can occur if its orbit is 
elliptical (or if it inspirals quickly),
as occurs with solar-system objects such as Io, Europa and Enceladus 
\cite[p. 171][]{md99}.
\cite{pcr79} give the heating rate as
\be
\frac{dE}{dt} = \frac{36\pi}{19} 
	\frac{e^2\rho^2}{Q\mu} 
  	\left(\frac{ R_a }{r}\right)^7 
	\frac{(G M_* )^{5/2}}{ r^{1/2}},  
\label{eq:tidal-heating}
\ee
where $\mu$ is the rigidity and $Q$ is the specific dissipation
rate (loss of energy per orbit). The implied radiative
equilibrium surface temperature is
\be
T_{\rm a, t} = 10^{-1.85} \, K\,
	\left( \frac{(e\rho)^2} {Q \mu_{12}}\right)^{1/4}
	R_2^{5/4} r_{10}^{-15/8},
\ee
where $\mu_{12} = 10^{-12}\mu$.  Tidal heating increases faster with
decreasing $r$ than other heating processes and could be quite large
for kilometer-sized objects that reach $10^8$~cm; however, tidal
disruption occurs before this takes place.  Thus, unless the rigidity
is small, tidal heating is at most secondary to the other heating
mechanims. 
 
To summarize, radiative heating of asteroids will dominate other
processes for very hot NS or those with large magnetospheric
luminosities beamed toward the asteroid belt.  For these objects,
entry into the magnetosphere will be prevented because material
is ionized outside the LC.    For other objects,
some combination of radiation and induction heating will evaporate
asteroids at radii of order $10^9$~cm, well inside the LCs of many objects.    
Thus we expect that objects that can accrete significant amounts of 
asteroidal material into their LCs will occupy certain regions
of the $P-\Pdot$ diagram that correspond to objects with different ages,
LC sizes, and ability to heat incoming objects.

For a fixed value of $T_a$ at the LC radius, a scaling
law for $\dot P$ vs. $P$ can be
calculated.  
Using the spindown age, $\tau_s = P/2\dot P$, as a proxy for
chronological age, we find $\dot P \propto P^5$ for magnetospheric
or induction heating and $\dot P \propto P^{1+1/2q}$ for heating from
thermal radiation whose temperature scales as $\tau_s^{-q}$.  
Cooling curves show a wide range of $q$, but for objects
$\sim 10^5$ yr old, $q\sim 0.14$ while at ages longer than 
about $10^6$ yr,  surface
temperatures fall off steeply, corresponding to $q\sim 0.4$ to 1
\cite[e.g. Figure 1 of ][]{yp04}.   
For a slow falloff with age,
$q\approx 0.14$, $\dot P \propto P^{4.6}$.   Thus 
heating by magnetospheric radiation or by thermal NS radiation
(in intermediate age objects) will yield
lines of constant asteroidal temperature
at the light cylinder distance from the NS that scale
roughly as $\dot P \propto P^5$, which are steeper than the putative 
inner-gap death line.  However, older pulsars will fall on a locus with
$\Pdot \propto P^{1.5}$ to $P^{2.3}$.


\begin{figure}[!ht]
\begin{center} 
\includegraphics[scale=0.45,angle=0]{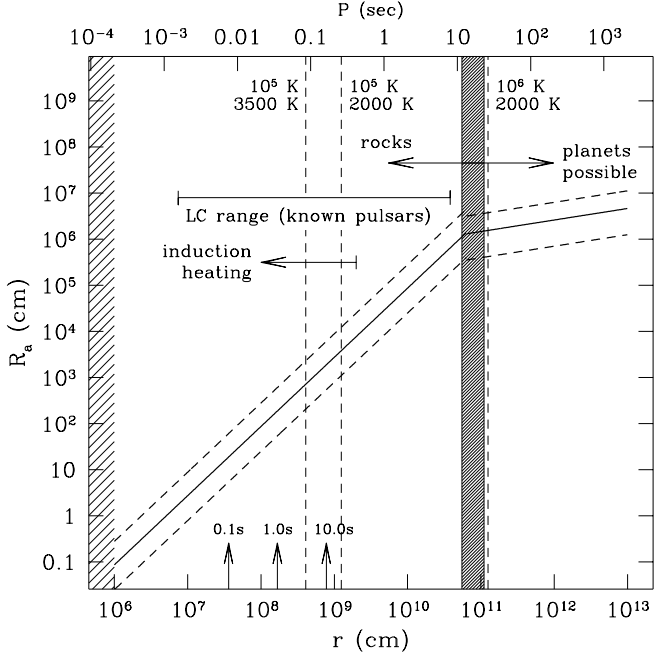}
\caption{\label{fig:Rvsr} 
Asteroid radius, $R_a$, plotted against distance
from the NS, $r$.  The top horizontal scale is
the period for which the light-cylinder radius $\rlc = cP/2\pi$
matches the radius on the bottom horizontal scale  and the horizontal
bar designates the range of LC radii for known pulsars.  The shaded
region at the left-hand edge designates the inside of the neutron star.  
For nominal parameters,
tidal disruption of gravitationally bound objects occurs
at $r_{\rm tg} \sim 10^{11}$~cm (vertical shaded region). 
At larger distances, 
planets may form while material
closer to the pulsar will only form rocks held together
by their tensile strength.  
The leftward going arrow indicates where induction heating starts
to be important for pulsars with canonical magnetic field strengths.
Slanted curves
to the left of $r_{\rm tg}$ 
designate rock sizes that can be held together by 
tensile strengths of $10^5$ and $10^8$~dyn~cm$^{-2}$ at densities
of 8 and 1 gm~cm$^{-3}$, respectively.  The 
slanted curves to the right  are the sizes of  planetesimals
that undergo runaway growth in an annulus of width twice their
Hills radius in the circumpulsar disk.  
Vertical dashed lines designate radii at which rocks will reach 2000 K for a
NS surface temperature of $10^6$~K and 2000 K or 3500 K at  $10^5$~K. 
Upward going arrows designate the radii at which an orbiting object
co-rotates with the NS for spin periods of 0.1, 1 and 10 $s$.
Parameters include  disk surface densities of 
$\Sigma = 1$ to 20 ${\rm g\,cm^{-2}}(r/10^{-2}\,{\rm AU})^{-3/2}$ 
for internal asteroid densities $\rho = 8$ to 1~g~cm$^{-3}$.   
}
\end{center}
\end{figure}

\section{Migration of Disk Material into the Magnetosphere}
	\label{sec:delivery}

Disks can supply inward moving asteroids, as we show here,
that stay neutral until
they enter the LCs of cooler and longer-period pulsars  
at  rates that are consistent with those for nulls and RRATs.
While inward migration from radiative effects
can occur rather slowly for a given object,
once it enters the magnetosphere, we assume that 
ionized material from the asteroids will be subjected to electric forces
that drastically alters its trajectory.  Alternatively, collisions in
the disk, which occur at interesting rates even for low-mass disks,
provide more direct injection trajectories into the magnetosphere.

Studies of accretion onto compact objects usually consider ionized gas,
which is subject to Lorentz forces that lead to the requirement that
material needs to orbit faster than the magnetosphere at the Alf\'ven
radius in order for it to be accreted.  
Slowly orbiting material is
flung outward by the propeller effect \cite[][]{is75}, resulting in spindown of
the star.  The situation is different if the material is neutral. 
An uncharged rock
(or one with small charge-to-mass ratio) will not respond to 
the magnetic field.  However, its surface charge and polarizability 
will lead to an interaction.  These effects are small for macroscopic (meter-sized)  objects.  The main consideration is the set of forces
on the rock as it evaporates and becomes ionized.   High-melting
temperature material can persist to radii of $10^{9.1}$~cm at 2000~K
and $10^{8.6}$~cm at 3500~K around a NS with surface temperature of 
$10^5$~K  (Figure~\ref{fig:Rvsr}).   These melting-point radii are
similar to those at which the rock will co-rotate with the NS,
as indicated for periods of 0.1, 1 and 10s in Figure~\ref{fig:Rvsr}.  
By the time the rock is ionized, which is probably outside
the corotation radius, 
$r_{co} = [GM_{\rm *} P^2/ (2\pi)^2 ]^{1/3}\approx10^{8.2}\,{\rm cm}\,P^{2/3}$,
for most pulsars, it encounters regions in the
magnetosphere where $\Epar\ne0$ and is electrically captured.

\subsection{Nominal Disk Parameters}
Consider an asteroid disk of total mass $M_a$ with 
minimum and maximum radii, $r_{1}, r_2 \gg r_1$, and
thickness $2H \equiv \epsilon_h r_{\rm 2}$ yielding a 
volume $V_d \approx (\pi/2) \epsilon_h r_{\rm 2}^3$, 
average surface density
\be
\overline\Sigma = \frac{M_a}{\pi (r_2^2 - r_1^2)}
\ee
and an optical depth 
\be
\tau = \frac{3}{4} \frac{\overline\Sigma}{\rho \tilde R_a},
\ee
where $\tilde R_a$ is a characteristic asteroid size and $\rho$ is its
internal mass density.   For a distribution of asteroid sizes,
$\tilde R_a = \langle R_a^3 \rangle / \langle R_a^2 \rangle$ which is,
for a power-law asteroid size distribution, 
$f(R_a) \propto R_a^{-m}, \, R_{a1} \le R_a \le R_{a2}$ with an
equilibrium distribution ($m=7/2$) \cite[][]{d69}, 
$\tilde R_a = \sqrt{R_{a1}R_{a2}}$. 
The mean time between collisions is
\be
t_c = \frac{2H}{\tau v_{\rm rms}},
\ee
where $v_{\rm rms}$ is the rms random velocity.    
The characteristic spreading
time for the disk due to collisions only is 
\cite[p. 496][]{md99} 
\be
t_{\rm coll} = \frac{1}{\tau \Omega_{\rm orb}} 
	\left( \frac{r_2 - r_1}{\tilde R_a} \right)^2.
\ee

For nominal parameter values 
$r_1 \ll r_2$,
$r_2 = 10^{-2}r_{2,-2}$~AU,
$\tilde R_a = 1\,R_{\rm a, km}$~km, 
and 
$M_a = 10^{-4}M_{a,-4}\,M_{\oplus}$,  
we have the following typical values for the asteroid disk:
\be
\overline\Sigma &=& 8.5\, {\rm g\,cm^{-2}}\,
	M_{a,-4} r_{2,-2}^{-2} \label{eq:Sigma}\\
\tau &=& 10^{-4.2}
	\left ( \rho \tilde R_{\rm a, km}\right)^{-1}
	M_{a,-4} r_{2,-2}^{-2} \label{eq:tau}\\
\Omega_{\rm orb} &=& 10^{-1.9}\,{\rm s^{-1}} 
	r_{10}^{-3/2} M_{*,1.4}^{1/2} \label{eq:Omega}\\
t_{\rm coll} &=& 10^{3.2}\,{\rm Gyr}\, 
		\rho \tilde R_{\rm a, km}^{-1} 
		r_{2,-2}^{11/2} 
		M_{\rm a, -4}^{-1} M_{*,1.4}^{-1/2} 
	\label{eq:lifetime}\\
t_c &=& 10^{2.9} \,{\rm yr}\, \epsilon_h 
	\rho v_{\rm rms}^{-1} r_{2,-2}^3 \tilde R_{\rm a, km}M_{\rm a, -4}^{-1}
	\label{eq:collisiontime}		\\  
\epsilon_h &=& \frac{2v_{\rm z, rms}} {\Omega_{\rm orb}r_2} 
	= 10^{-4.0} v_{\rm z, rms} M_{*,1.4}^{-1/2} r_{10}^{3/2} r_{2,-2}^{-1}
	\label{eq:epsilon}\\
N_a &=& 10^{8.2}\, 
	M_{\rm a, -4}\, \left (\rho \tilde R_{\rm a, km}^{3}\right)^{-1}.
	\label{eq:Na}
\ee
where $v_{\rm rms}$ and $v_{z, rms}$ are rms velocities in km s$^{-1}$,
with the z direction parallel to the disk's axis.  We have evaluated
$t_{\rm coll}$ using an orbital frequency corresponding
to an orbital radius, $r_2/2$.  
In these and other
calculations we ignore general relativistic corrections to orbital
parameters, though apsidal advance and Lense-Thirring precession will
be significant for orbits near or inside the LC.    However, these effects will likely be subdominant to the much stronger electromagnetic forces in the region of strong gravity (and strong electromagnetic field) close to the surfaces of the NS .  


\subsection{Debris Migration}
Collisions can direct some asteroids or their 
fragments into the magnetosphere.
However, additional processes now known to be significant in the solar
system are likely to also be important in the circumpulsar disk.
These include the diurnal and seasonal Yarkovsky effects for larger
rocks ($\sim 1$ m) and the Poynting-Robertson effect for small grains
\cite[e.g.][]{bls79}. 

The diurnal Yarkovsky effect results from the net force on an 
illuminated object whose
afternoon surface is hotter than its nighttime surface. It can lead
to an increase or decrease in semi-major axis depending on the lag 
angle of the hottest surface.   The seasonal Yarkovsky effect applies
to objects with spin axes tilted from their orbital plane, causing
a net force along the spin axis that always decreases the
semi-major orbital radius.   In the solar system the diurnal effect
is larger than the seasonal effect.  
Poynting-Robertson acceleration also 
decreases the orbital radius and is smaller than the Yarkovsky effect by
a factor $v_a/c$, all else being equal,  but is important for small grains 
that are isothermal for which the Yarkovsky effect vanishes. 
Near-Earth objects can migrate
up to $10^{-3}$~AU~Myr$^{-1}$ at 1 AU from the Sun 
from the seasonal Yarkovsky effect, which
we take as a fiducial value.  

The Poynting-Robertson effect drags small particles into the Sun on a time scale
$t_{\rm PR} = 4\pi\rho c^2  r^2 R / 3 L \approx 672\,{\rm yr}\, 
	\rho R_{-4} r_{\rm AU}^2 \Lsun / L$.
NS thermal radiation yields a time scale for grains 
of size $10^{-4}R_{-4}$~cm at $10^{10}r_{10}$~cm
\cite[e.g.][p. 153]{rl86}
\be
t_{\rm PR} = 16.8\,{\rm yr}\,
	\frac{\rho R_{-4}r_{10}^2}{R_{*,6}^2 T_{*,5}^4}.
\ee 
For nominal disk parameters, where disk-wide collisions are frequent,
a steady supply of grains will be injected
into the magnetosphere, similar to the interstellar flow proposed by
\cite{c85}.

We now focus on the seasonal Yarkovsky effect because it provides a viable
lower bound on the rate of  inflow into the LCs of
some pulsars. 
The Yarkovsky effect has been detected directly through 
measured accelerations of asteroids using radar 
techniques\cite[][]{ches03, nb04}.  
It plays a key role
in asteroid migration, especially to Earth-crossing orbits, by moving 
objects in the asteroid belt
into mean-motion resonances that then cause much faster orbital evolution.  
The seasonal Yarkovsky effect causes a secular
change in semi-major axis of the form 
\be
\dot r \propto  -\frac{(1-A)L}{c\rho R_{\rm a} \Omega_{\rm orb} r^2} 
	\left(\frac{\Delta T}{T}\right)
\ee
where $L$ is the luminosity, $\rho$ and $A$ are
the mass density and albedo of  the asteriod,
and $\Delta T/T$ is the fractional temperature difference caused by 
misalignment
of the spin axis from the normal to the orbital (or invariant) plane
(Rubincam 1998).  The above proportionality can be derived by assuming
simplistically that 
a spherical asteroid has a temperature enhancement $\Delta T$ for the 
hemisphere facing the direction of motion.
At $r= 1$~AU, $\dot r \approx -10^{-3}$~AU~Myr$^{-1}$ for a 
basaltic asteroid of 100 m size; 
values of $\dot r$ for iron asteroids are smaller  by about a factor of
two.   The effect asymptotes to zero for smaller asteroids as they trend
toward isothermality  and is smaller for more massive asteroids. 

The penetration depth for
thermal X-rays $\ell_X \sim 0.1-1\,\mu\, m$ \cite[][]{c85,v91} 
for 0.1 to 1 keV photons, so macroscopic objects will be
heated asymmetrically.  
The thermal skin depth  for iron with 
thermal conductivity
$\kappa = 4\times 10^6\,{\rm erg\,s^{-1}\,cm^{-1}\,K^{-1} }$, 
heat capacity 
$C_p = 5\times 10^6\, {\rm erg\, g^{-1}\, K^{-1} }$ and
density
$\rho = 8$~g cm$^{-3}$
is
\be
\ell_s = \left (\frac{\kappa}{\rho\Omega_{\rm orb} C_p}\right)^{1/2}
\approx 27\,{\rm cm} 
	\left(\frac{1.4\Msun}{M_{\rm *}} \right)^{1/4}
	r_{10}^{3/4}.
\ee  
Basalt and other materials have substantially smaller thermal conductivities 
yielding 3 times smaller skin depths,
so conceivable materials can sustain temperature gradients 
to distances near the LC if they are larger than a meter.   

We suppose that
rocks possess spins with random orientations owing to collisions or
chaotic dynamics, so any heating will drive the seasonal Yarkovsky 
effect.  
In a pulsar disk $\sim 10^{-3}$~AU from the NS, the Yarkovsky effect will
be larger than for the solar system
for fixed luminosity, but we must take into account 
both the thermal radiation from the NS and non-thermal radiation from
the magnetosphere.  
For a thermal luminosity
$L\approx 10^{28.9}\,{\rm erg\,s^{-1}}\, R_{\rm *, 6}^2 T_{\rm *, 5}^4$, 
the Yarkovsky drift rate will be 
\be  
\dot r \approx -10^{-7.2} \,{\rm AU\,Myr^{-1}} 
\left( \frac{\vert \dot r_{\rm ss, 1\,AU}\vert}{10^{-3}
	\,{\rm AU\,\,Myr^{-1}}}\right) 
	\frac{ R_{\rm *, 6}^2 T_{\rm *, 5}^4 } 
	     { r_{10}^{1/2} R_{\rm a, km} }
	, 
\ee
which is much smaller than the reference solar-system rate we have
adopted, 
$\vert \dot r_{\rm ss, 1\,AU}\vert\approx 10^{-3}
        \,{\rm AU\,\,Myr^{-1}}$,
owing to the small
thermal luminosity and the much larger orbital frequency.

Now we consider
the fraction of the pulsar's spindown energy loss rate, $\Edot$, 
that is in a form
that can heat a planetesimal.  The LC is at the transition
from the near zone to wave zone for rotational energy losses and most of
the spindown loss rate, $\Edot$, is in low-frequency waves that will 
have little effect on a rock's
dynamics.   Cheng (1985) concluded that relativistic
particles from the pulsar wind heat
grains negligibly while X-rays dominate the ionization of grains after
they enter the magnetosphere, a result that also applies to the
heating of grains and rocks in the current model.   
Typically the non-thermal X-ray luminosity of a pulsar
satisfies $L_X \approx 10^{-3} \Edot$, so 
\be
\dot r &\approx& - 10^{-5.5} 
	\,{\rm AU\,Myr^{-1}} \times \nonumber \\
	&&
	\left(\frac{\vert \dot r_{\rm ss,1\,AU}\vert }
		{10^{-3}\,{\rm AU\,\,Myr^{-1}}}\right)
        g_b r_{10}^{-1/2} R_{\rm a, km}^{-1}
	\left (\frac{L_X/\Edot}{10^{-3}} \right) 
	\left (\frac{\Edot}{\Lsun}\right),
\ee      
where we have used the beaming factor $g_b$ defined before, which can
exceed unity if the beam illuminates the asteroid but vanishes if not.
Thus the Yarkovsky drift rate can be much larger for nonthermal radiation
than for thermal radiation.

\subsection{Asteroid flux into the light cylinder}

\subsubsection{Yarkovsky Migration}

The flux of asteroids through a cylinder centered on the NS with fiducial
radius $r$  and surface area 
$A_c = 2\pi \epsilon_h  r_{\rm 2}  r$ is
$\dot N_a  = n_a A_c \vert \dot r \vert \approx 2 N_{\rm a} r \dot r r_2^{-2}$,
or 
\be
\dot N_a &\approx& \frac{3}{2\pi} \frac{M_a}{\rho \tilde R_a^3} 
	\frac{r \vert \dot r\vert}{r_2^2}
	 \approx \frac{3}{\pi} \overline\Sigma 
	\frac{r\vert \dot r\vert}{\rho \tilde R_{\rm a}^3} \\
	&\approx&
	10^{-11.5}\,{\rm s^{-1}} \times \nonumber \\ 
	&&
	\frac { r_{10}^{1/2} R_{*,6}^2 T_{*,5}^4 } 
	      { \rho \tilde R_{\rm a, km}^{4} } 
	\left (\frac{\overline\Sigma}{8.5\,{\rm g~cm^{-2}}} \right)
 	\left(\frac{\vert \dot r_{\rm ss,1\,AU}\vert}{10^{-3}\,{\rm AU\,\,Myr^{-1}}}\right)
\ee
A rate $\dot N_a \sim 10^{-3} - 1$ s$^{-1}$ is probably sufficient 
to account for nulling, so asteroids of 10 m size around NS slightly hotter
than $10^5$~K or a disk more massive than $10^{-4}\Mearth$ can
bring the rate into this range.  

The lifetime of the disk to inward migration is
\be
\tau_d &\approx& \frac{N_a}{\dot N_a} = 
	\frac{r_2^2 }{2 r \vert \dot r\vert} 
	\nonumber \\ 
	&\approx&	
	10^{3.1} \,{\rm Gyr}\, 
	\frac{ r_{2,-2}^2 R_{\rm a, km}}
	{ R_{*,6}^{2} T_{*,5}^{4} r_{10}^{1/2} 
	} 
	\left(\frac{10^{-3}\, {\rm AU\,Myr^{-1}}}{\dot r} \right),
\ee
about ten times shorter than the collisional lifetime for nominal parameters.
However, for parameter  values that make the influx rate $\dot N_{\rm a}$
large, the disk lifetime will be substantially smaller, but still  larger
than the characteristic lifetime of a canonical radio pulsar. 

\subsubsection{Collisional Migration}

A debris disk surrounding a neutron star has many characteristics similar 
to a planetary ring system and therefore many of the results of ring 
dynamics can be applied to this discussion.  Inelastic 
collisions between particles determine the structure and evolution of a disk.  
On short time scales ($\sim \tau^{-1}$ orbits, where $\tau$ is the 
optical depth of the disk) these collisions will result in the asteroids 
moving towards circular, coplanar orbits.  On much longer timescales, 
these collisions widen the disk, with most of the 
mass transported inwards and most of the angular momentum transferred 
outwards.  

\cite{1984Stewart}  calculate the inward flux of material 
$\dot{m}$ in a viscous disk.  This hydrodynamical result is consistent with 
a result from the more general approach in which the asteroid behavior is 
stastically modelled using the Boltzmann equation, presented in the 
same paper.  They assumed that the asteroids are identical, indestructable, 
nonspinning spheres, which have a size much smaller than the 
inter-asteroidal spacing.  Furthermore, they assume that the velocity 
dispersion of the particles has a Gaussian form, consistent with simulations 
of ring dynamics.   The total flux $\dot{N_a}$ through radius r is then,

\begin{equation}
\dot{N_a} = \frac{9}{2} \frac{1} {\rho R_a^3} \frac{1}{\Omega r}\frac{\partial}{\partial r} ( r^2 \Omega \Sigma \nu),
\label{collision_flux}
\end{equation}
where $\Omega$ is the Keplerian angular frequency and $\Sigma$ is the surface density.  The viscosity $\nu$ was calculated numerically by \cite{1978Icarus.34..22}, by balancing the energy dissipated during the inelastic collisions and the energy genertion due to the viscous stress.   

\begin{equation}
\nu \approx 0.15 \frac{v_{\rm rms}^2}{\Omega} \frac{\tau}{1+\tau^2}.
\end{equation}   
 
Assuming that $\Sigma$, the velocity dispersion $v_{\rm rms}$, 
and the optical depth $\tau$ do not vary greatly through the disk, 
the influx rate at the inner edge of the disk 
(set to be at the light cylinder) can be estimated:
\begin{equation}
\dot{N_a} = 10^{6.5} {\rm s}^{-1} \frac{\tau}{1+\tau^2} r_{2,-2}^{-2} R_{\rm a,2}^{-3} M_{\rm a,-4} M_{*,1.4}^{-\frac{1}{2}} \rho^{-1} v_{\rm rms}^2  P^{\frac{3}{2}},
\end{equation}
with $P$ in seconds, $\rho$ in g cm$^{-3}$, and  $v_{\rm rms}$ in km s$^{-1}$.  
Substituting in the previous expression for optical depth, and assuming it 
is much less than unity, we obtain
\begin{equation}
\dot{N_a} = 10^{5.3} {\rm s}^{-1} r_{2,-2}^{-4}R_{\rm a,2}^{-4} 
M_{\rm a,-4}^{-2} M_{*,1.4}^{-\frac{1}{2}} 
\rho^{-2} v_{\rm rms}^2 P^{\frac{3}{2}}, 
\end{equation}
This migration rate is large and would provide material at a rate 
sufficient to disrupt emission processes, even if the mechanism 
for injecting material into the magnetosphere is inefficient.



\section{Inside the Magnetosphere}\label{sec:magnetosphere}
The standard picture is that coherent radio emission in pulsars requires 
counterstreaming particle flows that include $e^{\pm}$ pairs produced
in cascades. Sustained cascades require acceleration regions in the
magnetosphere that yield particles energetic enough to radiate gamma-rays
that then pair annihilate. As pulsars spin down, cascades cease as
the electric fields in acceleration region(s) lessen.  However, 
dormant regions can produce triggered pair avalanches that cause
coherent radio transients.   

\subsection{Asteroid Evaporation}

\begin{figure}[!ht]
\begin{center}
\includegraphics[scale=0.45, angle=0]{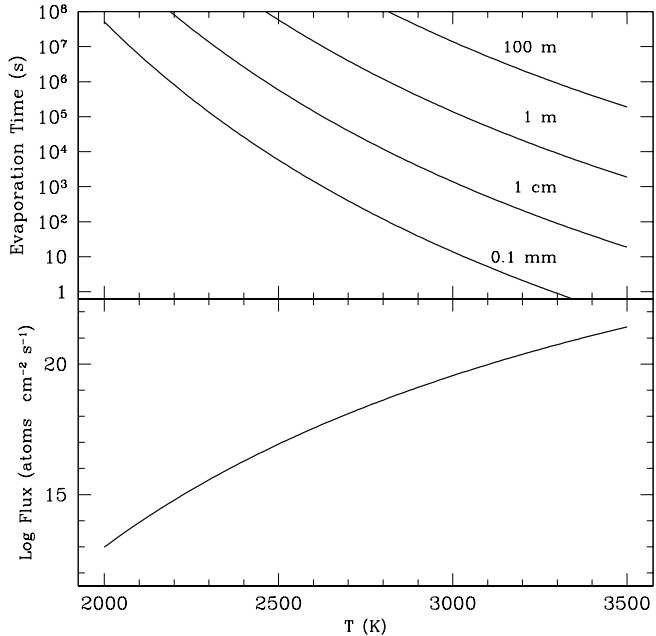}
\caption{\label{fig:evap12}
Evaporation of material from hot asteroids assuming an evaporation
pressure for carbon using evaporation pressure data quoted from
\cite{c85}.
Top: Evaporation times for different sizes. 
Bottom: Evaporation flux. 
}
\end{center}
\end{figure} 

\begin{figure}[!ht]
\begin{center}
\includegraphics[scale=0.45, angle=0]{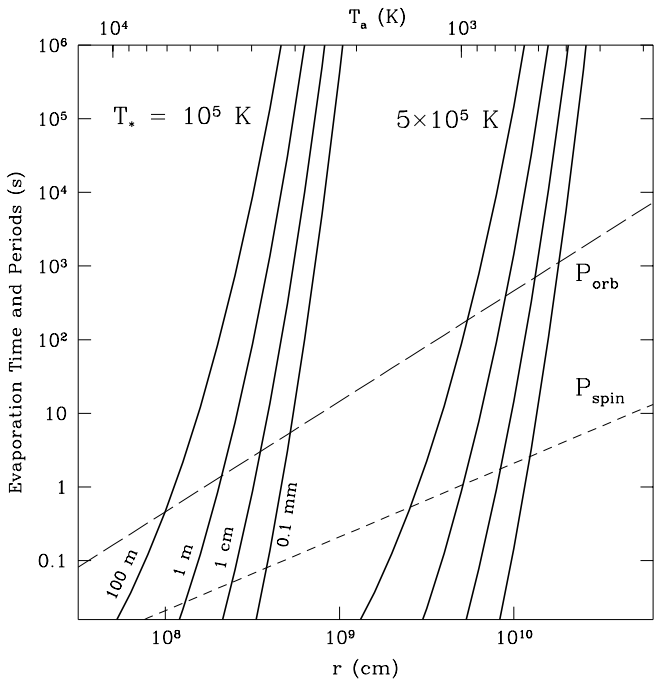}
\caption{\label{fig:evap_vs_r}
Evaporation times for asteroids of different sizes for two
different values of NS surface temperature, as labelled.
Also shown (dashed lines) are the Keplerian orbital period
and the spin period, the latter assuming that $r$ corresponds
to the light cylinder radius. 
}
\end{center}
\end{figure} 

Asteroids rapidly evaporate when they reach radii below $10^{10}$~cm
for $T_* = 5\times 10^5$~K and below $10^9$~cm for $T_* = 10^5$~K.
Figure~\ref{fig:evap12} shows evaporation times for objects
of different size and evaporation fluxes
assuming carbon composition \cite[][]{c85}.
Figure~\ref{fig:evap_vs_r} shows evaporation times vs. distance
from the star, assuming that the asteroid surface is in equilibrium
with surface radiation from blackbody radiation.   
These figures show that evaporation times are comparable to the 
orbital period only after asteroids are deep within the magnetospheres
of long-period pulsars.  Also, for larger objects ($\sim 100$~m),
evaporation rates are comparable to the Goldreich-Julian particle rate, 
$\dot N_{\rm GJ}$,  well before they are completely evaporated.    
These results confirm the basic picture that asteroid evaporation
occurs at locations and with rates suitable for modification
of magnetospheric electrodynamics.  Moreover, the range of asteriod
sizes expected for a highly collisional asteriod belt  can account
for the broad range of time scales seen in coherent radio emission. 
The curves shown in Fig.~\ref{fig:evap_vs_r} do not include the effects
of ohmic heating (mentioned earlier).  Ohmic heating 
from ${\bf  v \times B}$ currents 
may be significant (see Appendix) 
during the final stages of evaporation and cause the evaporation times
to vary more steeply with radius than shown in the figure.  The relative
importance of the various heating effects will be strongly object
dependent, so we defer a detailed comparison to a future analysis. 

\subsection{Electromagnetic Capture}

Ionized evaporant from an inspiraling or infalling asteroid is rapidly
spun up to the NS's spin rate, $\Omega_*$ owing to the typically small
electron gyro-radius, 
$r_g \approx 10^{2.3}\,{\rm cm}\, \left (r/R_{\rm LC} \right)^4 P^3$
(without relativistic corrections and ignoring asteroidal motion). 
The detailed trajectory of the material depends on its field-parallel
velocity prior to ionization (e.g. collisional motion) and on any
parallel electric fields.   Because the lifetime for many asteroids
exceeds a spin period, an asteriod and its evaporant
will encounter gaps.  Ionized material is accelerated
inside such gaps, contributing to the current flow in the magnetosphere,
some of which is subsequently lost to flows outside the LC.   The fate
of material captured in the closed magnetosphere  is not clear.  It can
remain tied to magnetic field lines but some may diffuse to larger radii
or the open field-line region from which it is accelerated to
the NS or lost by the magnetosphere.

\subsection{Currents and Gaps}

Pulsar magnetospheres will adjust the charge
density to $\ngj$ in order to become force free (i.e.~vanishing
electric field parallel to $B$, $\Epar = 0$).  But 
boundary conditions (or, as formerly assumed,  
large work functions at the NS surface)  
 prevent this in ``gaps'' where $\Epar \ne 0$.   
Gap sites include the polar cap above the NS surface in the open
field line region \cite[][]{rs75}, the slot gap at the interface between
the open and closed field-line regions 
\cite[e.g.][]{saf78, a83, mh03}
and the outer gap that bounds the null surface
$\Omega\cdot B(r) = 0$ \cite[e.g.][]{chr86, ry95, crz00, hhs03, zcjl04, t+06}.

Particle acceleration in gaps can yield copious gamma-ray emission that 
drives electron-positron ($e^{\pm}$) pair cascades  if there is sufficient
residual field $\Epar\ne 0$, which 
will depend on other sources of charged particles,
such as thermionic emission from the NS surface or injection 
of charges from external sources.

We assume that the standard picture
is at least approximately correct  because 
(a) acceleration of particles to high energies in the open field-line region 
is clearly needed to produce beamed radio emission that is
consistent with the observed widths and polarization of pulses;
(b) subpulse drift behavior is qualitatively consistent with
\ExB\ drifts around the magnetic axis, which rely on the existence
a gap field; and
(c) if surface magnetic field estimates are correct, the conditions
exist for pair production to occur. \cite{j+01} discuss an 
alternative to this view.   Whether pair production is
in the form of sustained cascades  or in episodic avalanches is open to
question for particular objects, with the answer dependent on
NS properties and on external triggers.    
Since the accelerating field
$\Epar \propto \rho_{\rm e}(r) - \rhogj(r)$ in the corotating frame,
 any excess particles
injected from outside the LC that diminish the charge density difference
(or change its sign) must drastically alter the population of particles and
the fraction of $\ngj$ needed to do so can be small. 
Drifting subpulse patterns are sometimes coherent, other times disrupted,
suggesting that the \ExB\ drift rate is time variable.  Our model provides
a natural mechanism for this variability.

\subsection{Coherent Radio Emission}

In our model, neutral
particles enter the magnetosphere as rocks or grains, dissociate and ionize
from thermal and perhaps magnetospheric radiation, and perturb the
current flow.  In a pulsar with an active cascade, the intruding
charges can reduce $\Epar$ and thus the density of cascade $e^{\pm}$ pairs.
Presumably this reduces the altitude of radio emission, which is tied
to one of the plasma frequencies defined in the system, or completely detunes
the coherent amplification by altering the growth rate of the plasma
instability that  provides the coherence.
As NS spin down, cascades cannot be sustained and
gap accelerators become quiescent. Charges from accreted rocks may 
temporarily re-activate the $e^{\pm}$ cascade in one or more of the gaps,
leading to a burst of coherent radio emission.

Enough uncertainty exists
about the structure of these gaps that we sidestep any detailed modeling
of the process by which deconstructed asteroids disrupt the electrodynamics.
Disruption undoubtedly depends on NS properties such as the surface
temperature and spin rate.  It also depends on the trajectories of incoming
asteroids, whether they slowly spiral in to regions where
$\Epar\ne0$, or alternatively, are on highly eccentric orbits
arising from collisions or resonances outside the magnetosphere.   

There appear to be ample
configurations and mechanisms by which disruption can occur, including
(a) direct injection of charges into the polar-cap gap and slot gap;
(b) perturbation of the outer gap by itself; and 
(c) direct injection into the outer gap, which triggers an inward
moving  cascade that perturbs the inner gap region(s). 
We consider the most likely case to involve contamination of the outer
gap, primarily because it borders the LC and its volume
presents a large target for incoming material \cite[e.g.][]{t+04,t+06}. 
In long-period, cool pulsars, the outer gap is unable to sustain
pair cascades that involve pair production from gap 
gamma-rays interacting with X-rays from the NS surface.  However, 
an external flux of particles can trigger an avalanche that, in turn,
may perturb the inner-gap region of the magnetosphere.  Such perturbation
could enhance the pair production in some cases or it could quench
the pair production in others.  Consequently, perturbation of the outer
gap may lead to a burst or a cessation (i.e. a null) of radio emission.

Perturbation of a gap region by injected charges may increase pair
production in some objects and decrease it in others, depending on
whether the region is ``starved'' of particles and on the potential
drop across it.  In active radio pulsars with a steady $e^{\pm}$ cascade
sustained by a slight deficit from the GJ charge density, injected 
charges will diminish $\Epar$, thus reducing the energies of accelerated
particles and the number of radiated gamma rays that can pair produce. 
The pulsar remains radio loud if the conditions are sustained for driving
a plasma instability that is responsible for the coherence and if coherence
occurs at detectable radio frequencies.   A change in radio emission altitude
is expected and it could be lowered if it depends solely on plasma
densities or it might increase if the instability growth rate becomes longer. 
In objects that are mostly radio-inactive because they cannot sustain
pair cascades,  injected charges can provide the seed particles needed to
initiate a short-lived cascade.

External particles need not provide 
the full GJ current in order to perturb the radio emission, so the
actual mass needed to produce observable effects is probably orders of 
magnitude smaller than given in \S\ref{sec:obscons}.   
In fact, if the GJ charge density were variable by
100\%, large torque changes would be produced.  Depending on the time
scales of such variability, torque fluctuations would be manifested as an observable contribution to the ubiquitous timing noise found in pulsars.

\subsection{Emission Altitude and Beaming}\label{sec:altitudes}

Most pulsar models invoke the two-stream instability as the source
of plasma bunching that underlies the strong coherence of the radio
emission \cite[e.g.][]{rs75, u87, am98}.  Models differ in the nature of
the two streams but assume that 
the instability's growth time must be smaller than the propagation time
from the NS surface, from which the plasma originates and propagates at $c$.
This requirement is highly restrictive and leads to
the proposal \cite[][]{u87, am98} that variability near
the NS surface causes particle clouds to overtake those 
from earlier discharges
and leads to suitable instability growth times.   When particles are injected
into the magnetosphere from asteriods, they provide an additional stream 
of particles that can enlargen the parameter space for producing
coherent radio emission.

In the two-stream instability,  the maximum growth rate is at a frequency
proportional to the local plasma frequency, which is a function of radius, 
$\omega_p \propto \left (\npm/\gamma \right)^{1/2}$, and which also depends on 
relevant Lorentz factors for the particle streams.  
The equation for a dipolar field line, 
$r(\theta) = R \sin^2\theta / \sin^2\theta_0,$
where $\theta_0$ is the angle of a field line at the NS surface,
implies that emission from relativistic particles
is beamed in directions parallel to the field line tangents that make
an angle $\theta/2$ from the radial vector.  Emission is therefore
beamed into polar angles $\theta_e = 3\theta/2$, for small angles near
the magnetic axis and well inside the LC. 
If the Lorentz factor remains constant, a change in pair 
density induces corresponding changes in emission altitudes and angles,
\be
\frac{\delta\npm}{\npm} =
-3\frac{\delta r}{r} = -6 \frac{\delta \theta_e}{\theta_e} 
\ee
where the minus signs signify that an increase in pair density  induces
a reduction in emission altitude and emission angle. 
Asteroidal injection may enhance or diminish the pair density,
so both increases and decreases can be expected. 
A given object may show a change in pulse profile (mode change), 
become too weak to detect because its beam narrows and thus misses
the line of sight, or become detectable as a result of a wider beam.    

\subsection{Subpulse Drift Rates}

Modulation of the charge density by injected charges alters not
only $\Epar$ but also the \ExB\ drift of structures within a gap region,
such as those offered as explanations for the drifting subpulse phenomenon.
\cite{rs75} derive a drift velocity for the polar-cap gap region
\be
v_d 	\approx \frac{c\Delta \Phi}{B_s R_*^{3/2}} 
		\left(\frac{cP}{2\pi} \right)^{1/2}
	\approx 10^{3.8} \,{\rm cm\,s^{-1}} P^{1/2} B_{12}^{-1} \Delta\Phi_{12}, 
\ee
where $\Delta\Phi = 10^{12}\Delta\Phi_{12}$~volts is the typical 
potential drop along the field lines in the gap.   Injected charges
will alter this potential drop, so the drift velocity will be
affected proportionally. 

\subsection{Spindown Energy Loss}

\cite{c85} writes the total spindown luminosity as
\be
\dot E_{\rm total} = \dot E_{\rm rad} + 
	\frac{\Omega^4 B^2 R^6}{c^3} \left(\frac{I_*}{I_{\rm GJ}}\right),
\ee
where the first term is the contribution from low-frequency dipole radiation
and the second term results from the torque exerted by magnetospheric
currents; $I_*$ is the actual current and
$I_{\rm GJ}$ is the Goldreich-Julian current.   Cheng
considered interstellar grains that intrude into the LC and contribute
to $I_*$.   For material originating from a circumpulsar disk, the effects
are largely the same, but the delivery of extrinsic
material and its likely episodic behavior are different.  
Intrinsic contributions
to $I_*$ include ions and $e^{\pm}$ pairs.  Without an extrinsic component,
the current may be steady with shot-pulse modulations on the
light-travel time across accelerating gaps ($\ll P$) or \cite[][]{cr80, j82}
it may show cyclical behavior on thermal time scales for the polar cap,
which Jones (1982) estimates to be several thousand seconds.    By introducing
an extrinsic contribution that has a wide range of time scales,
the rich variety of radio intensity variations 
--- and their variation from object to object --- can be explained. 

\subsection{Torque Fluctuations and Timing Noise}

As discussed earlier, TOA variations will result from the 
recoil of the NS in response to an ensemble of orbiting asteriods. 
Accreting asteroids will induce true spin fluctuations of the NS through two
mechanisms.  First, spinup of a mass $m_a$ through electrical capture
in the magnetosphere implies corresponding spindown of the NS.
Second, an increase in current from injected charges will produce a 
torque pulse that acts on the  NS.  We consider these in turn.

Suppose that an effective fraction 
$f_m$ of the mass $m_a$ of an asteroid is ultimately
spun up into corotation from an orbit that is much slower than the NS
spin.  Once ionized, ions and electrons will be accelerated in opposite
directions, so the spunup mass will be significant only for cases where ions
are accelerated inward.   When ions are accelerated outward and 
lost from the magnetosphere, a torque
will be applied to the NS if the net acceleration is non-radial, as is
likely.   Thus, we expect in either case that the effective value of 
$f_m$ may be sizable.  The change in NS spin rate for an electrical
capture radius, $r_c$, is 
\be
\Delta\Omega_* = -(\Omega_* m_a r_c^2/I_*) (1 - \Omega_K / \Omega_*),
\ee
where $I_*$ is the NS's moment of inertia and $\Omega_K$ is the Keplerian
angular velocity, which we assume is much less than $\Omega_*$
because objects evaporate well outside the corotation radius.
The resulting random walk
in spin rate will have a strength parameter \cite[e.g.][]{cg81}
\be
S_{\rm FN} &=& {\cal R} (\Delta\Omega_*/2\pi)^2 
	= 
	{\cal R} (f_m m_a r_c^2/I_*)^2 (\Omega_*/2\pi)^2 
	\nonumber \\ 
	&=& 10^{-25.3} \,{\rm s^{-3}}\,
	\frac{\cal R}{10\,\rm yr^{-1}} 
	\left(f_m \rho R_{\rm a, km}^3 r_{c,10}^2 / I_{45} P^2\right)^2,
\ee
where ${\cal R}$ is the event rate, assumed equal to the
asteroid injection rate, and 
parameter values were chosen to yield strength parameters 
and rms pulse phases similar to
those inferred from pulsars \cite[e.g.][]{ch80, cd85, a+94, h+04b}.
The rms pulse phase variation over a time span $T$ is 
\be
\sigma_{\phi}(T) &=& \left ( \frac{S_{\rm FN} T^3}{12} \right)^{1/2}
	\nonumber \\
        &\approx& 
	10^{-1.9}\,{\rm cy}\times \nonumber \\
	&& \left(\frac{\cal R}{10\,\rm yr^{-1}}\right)^{1/2}
        \left(\frac{f_m \rho R_{\rm a, km}^3 r_{c,10}^2  T_{\rm yr}^{3/2}}
	{I_{45} P^2}\right),
\ee
which is comparable to measured rms timing noise variations for
nominal parameter values if the effective fraction $f_m$ is not small.

The second mechanism, fluctuations in current,
 will alter the torque on a NS to produce
spin noise. \cite{cg81} considered a superposition of 
torque pulses associated with nulling and  
modeled the resulting timing noise, which also would  be
a random walk in the spin frequency.
Episodic accretion events that produce torque pulses of duration
$W$ by a fractional change in current, 
$f_I = \delta I_*/I_*$, yield  steps in spin frequency,
$\Delta \Omega_* = f_I \dot \Omega W$. 
The strength parameter for the random walk, 
$S_{\rm FN} = {\cal R} W^2 (f_I \dot \Omega / 2\pi )^2$,
corresponds to an rms spin-phase (for $W = 10^3 W_3$~s) 
\be
\sigma_{\phi}(T)  
	\approx 
	10^{-2.8}\,{\rm cy}\, f_I 
	\left( {\cal R} W \right)^{1/2}  
	W_3^{1/2}
	\Pdot_{-15}P^{-2} T_{\rm yr}^{3/2},
\ee
also comparable to measured timing noise.
Asteroidal intrusion into the LC is
more likely in older, long-period pulsars with small values of $\dot P$
for which $\sigma_{\phi}$ is small, whereas measured timing noise is
larger for short-period objects with large $\Pdot$.   It seems likely
that there are several processes that contribute to timing noise,
with extrinsic current sources playing a major role only in some objects. 
The recent discovery of 50\% increases in $\Pdot$ in pulsar
B1931+24 \cite[][]{k+06} with $W \approx 10$~d, ${\cal R}W \sim 0.2$,
and $f_I \sim 1$ is consistent with this picture for timing noise,
albeit for a case where the individual torque events are easily seen
in the timing data.  For pulsars with faster rates and shorter
event durations, $W$, the torque events may be inferred only
statistically rather than through individual changes in $\Pdot$. 

The ratio of spin-frequency perturbations for the two cases
is
\be
\frac{\Omega_*^{(2)}} {\Omega_*^{(1)}} = 
	\frac{ f_I W I_*\Pdot} {f_m m_a r_c^2 P} 
	\approx 10^{-2} 
	\frac{ f_I W_3 I_{45}\Pdot_{-15}}{f_m m_{a,15} r_{c,10}^2 P}.
\ee
Spindown from mass loading dominates for the rather large effective asteriod
masses assumed ($f_m m_a = 10^{15}f_m m_{a,15}\,{\rm g}$), and the large
capture radius; adjustment of these and other parameters can modify the
ratio by several orders of magnitude either larger or smaller. 

Asteroid-induced torques may also induce changes in 
spin-axis orientation,  which would amplify their effects on the 
measured pulse phase.  The change in orientation will depend on
how the instantaneous current density deviates from the average current
density, so is hard to predict.   If the mean torque depends
on the angle $\alpha$ between the spin and magnetic axes as
$f_{\alpha}$, then the induced change in $\dot P$ will be 
\cite[][]{c93} 
$\delta \Pdot \sim \delta\theta \Pdot (\partial f_{\alpha}/\partial\alpha)$.
Even very small angular changes can produce easily detectable, accrued
pulse-phase changes from this effect.

\section{Applications} \label{sec:applications}
\subsection{Nulls,  Drifts and Mode Changes}\label{sec:nulling}

Charges injected from asteroids may 
be sufficent in number that they reduce or shut off the pair cascade and
thus alter or extinguish the radio  emission.  Several cases may apply.
First, injected charges may directly perturb the flow in the gap region
that is responsible for producing the radio-emitting particle flow;
it is commonly assumed that this is the polar-cap gap at the NS surface.
In this case, we would also expect the current to increase during
a null, leading to an increase in spindown torque, as described above.   
Alternatively,  asteroid charges  are injected
into an  outer-gap region, producing an avalanche of pair production with
one charge sign flowing into the inner gap.   

All things considered, we expect nulls to be most prominent in 
cool, long-period pulsars for which a wide range of asteroidal conditions
can yield influx into their large LCs.   
Nulls are not impossible in younger, hotter objects, but the requirements are
more selective on asteroid size and on trajectories into the LC.
From a given orbital radius, the LC 
is a fraction $\sim (cP/4\pi r)^2\ll 1$ of the sphere, so collision
fragments large enough to avoid evaporation outside the LC must be on
highly restricted trajectories  to hit the target. 

Pulsars that show nulling often also show drifting subpulses with
variable drift rates and changes in pulse profile (mode changes).
In some cases, drift rates jump between two or more preferred
values.    We interpret such objects as ones that can sustain current flows
and pair-production cascades, on average, but that are modulated by
accreted asteroids according to the processes we have identified.  
Injected charges will alter
the local difference between the actual and Goldreich-Julian charge
and current densities, $\rho - \rho_{\rm GJ}$, which will change
the parallel electric field, $\Epar$.  Boundary conditions at the interfaces
between the gap, the closed-field-line region, and the NS surface then
imply changes in the transverse electric field and, hence, in the
\ExB\ drift rate. Injected charges therefore can account for variations
in drift rate  and it is not surprising to find objects that show
both nulls and variable drift rates. 
Why should drift rates have preferred values?  This is equivalent to
asking, why should there be quantized values of the parallel
electric field?   We offer no definitive answer, but it seems possible
that particle source regions in the polar-cap region, which drift
and themselves appear finite in number \cite[e.g.][]{dr99}, may appear
and disappear as discrete entities.  Alternatively, if the current and
particle flows consist of several particle types, with one or  more
 depending on injection from external sources,  the current and potential
drop may show discrete states. Finally \cite[][]{g+05} propose that
subpulse drift is associated with drift waves in the open field-line region
of the magnetosphere that modulate the particle accleration. 
Asteroidal influx, in turn, would modulate this modulation.

\subsection{A Cue-ball Model for Quasi-periodic Bursts and Nulls  in B1931+24}\label{sec:B1931+24}

The long $\sim 40$~d quasi-periodic bursts from B1931+24 with about 20\% duty
cycle represent a time scale that is very hard to understand from processes
internal to the magnetosphere or the NS.  Periods of about this
duration are associated
with Tkachenko oscillations of superfluid vortices \cite[][]{cesc03}, but presumably
these are cyclical with high oscillation ``Q'' from which it is not
clear how low-duty cycle bursts would be produced. Orbital
processes seem richer in possibilities and thus more promising. 

B1931+24 itself is a typical canonical pulsar 
($P = 0.81$,  average $\langle \Pdot \rangle = 10^{-14.1}$, and spindown age
$\tau_s = P / 2\langle \Pdot \rangle = 1.6$~Myr)
yet the effects seen from it 
(bursts and nulls, a 50\% increase in $\Pdot$ during bursts, and
no short nulls during a burst) are uncommon  among known objects. 
Observational incompleteness due to
the impracticality of continuous sampling of large
regions of the sky may play a role, but the observed effects
are probably physically rare.
Therefore we seek special circumstances to explain the
observed phenomena.  

\subsubsection{Eccentric Asteroid Model}

In the context of our model, the burst period and
duty cycle are naturally explained with a large asteroid of radius $R_b$
 moving in an eccentric 40-d orbit with
semi-major axis $a = 0.26\,{\rm AU}\, (\Porb / 40\,{\rm d})^{2/3}$.
At this distance, $R_b$ can be as large as 100~km (c.f. Figure~\ref{fig:Rvsr}). 
This ``cue-ball'' or ``mini-Nemesis'' asteroid\footnote{Nemesis is the
name of the putative solar companion star that was hypothesized to be
the cause for quasi-periodic, terrestrial extinction events with period
$\sim 36$~Myr due to perturbation of the Oort cloud of comets \cite[][]{r86}.}
 induces observable effects
by deflecting smaller asteroids into the loss-cone
for light-cylinder insertion in much the same way
that solar-passing stars perturb comets in the Oort cloud.  Injection into
the magnetosphere would be stochastic, so the observed
continuity of the bursts,
once they commenced, suggests efficient injection of a large number
of objects.    Burst durations also would be stochastic.  
We consider implausible a model where the cue-ball asteroid directly injects
charges into the magnetosphere.  To do so requires an
eccentricity very near unity in order that the orbit's minimum radius
$a(1-e)$ be comparable to the LC radius.   Tidal heating 
(c.f. Eq.~\ref{eq:tidal-heating}) is also unlikely to be significant
for an asteroid with 40~d orbital period with reasonable eccentricity. 

During short duty cycle bursts, the spindown
rate increases, suggesting that the asteroid induces enhancements to
the current flow.
We associate such episodes to the periastron of a large asteroid
because, in an elliptical orbit, an object spends less 
time nearer than farther from the star.
For example
in an orbit with $e=0.61$, the cue-ball asteroid spends 20\% of its
time at radii less than twice the minimum NS distance of $a(1-e)$.    
If only a 10\% increase in distance is sufficient to terminate the bursts,
an eccentricity of 0.25 is implied.   The opposite case, where
bursts are associated with apastron and a termination of asteroidal
influx, would imply a lower-eccentricity
orbit but does not readily explain the increase in $\dot P$ during bursts.

For a large asteriod in the densest part of the disk of small asteroids,  
the mean time between collisions is
(c.f. Eq.~\ref{eq:collisiontime}) 
\be
t_c = 10^{7.4}\,{\rm s}\, 
		\frac{\epsilon_h\rho}{\Delta v}
		\left( \frac{R_a}{R_b} \right)^2
		R_{\rm a,2} r_{2,-2}^3 M_{a,-4}^{-1},
\ee
where $\Delta v$ is the relative velocity.  This velocity difference is, 
for small asteroids in circular orbits of size equal to the periastron 
distance of the cue-ball asteriod, 
$\Delta v = \sqrt{(1+e)^{1/2}-1)(GM/r)} \sim (e/2)v$, a potentially
large velocity.
For nominal parameters (Eq.~\ref{eq:Sigma}-\ref{eq:Na}) 
$t_c \sim 1$~s for $R_b/R_a = 10^3$ (i.e. $R_b = 1$~km), implying a large
number of collisions per orbit,  which may imply that the cue-ball asteriod's
orbit will evolve too quickly. 
Thus, either the periastron
of the cue-ball asteriod is outside the densest part of the asteroid disk
or the disk is much less dense than our nominal disk.

The eccentricity of a cue-ball asteriod suggests that the object is
part of a family of large objects that co-exists outside the small
asteriod belt of objects that are injected into the pulsar's magnetosphere.
This follows from the need to account for the cue-ball's eccentricity.
In the solar system, near-Earth objects (NEOs) exist as a result of
chaotic dynamics in main-belt asteroids induced by resonances with
Jupiter and possibly Mars \cite[][]{md99}.    Thus there may be additional large
objects orbiting B1931+24.  Given the large change in $\Pdot$ in
the on state for the pulsar, it is possible that orbital recoil
effects on the TOAs of the pulsar are masked.  Thus it is of interest
to carefully analyze timing data to search for the individual effects of 
multiple orbiting objects. 

\subsection{Orbital and Disk Lifetimes}



Requirements on the orbital collision model are that
(a) the cue-ball orbit must be stable for a long period of time, $t_b$,
that we take to be $t_b = 10^5 t_{\rm b,5}$~yr. 
and (b) the influx into the LC of the pulsar must be at a rate that
can yield continuous and sustained bursting over a $\sim 8$-day period. 
The first condition places a constraint on the change in orbital momentum
per orbital period based on collisions occuring at a
rate $t_c^{-1}$ for a fraction of the orbital period, $f_b P_{\rm orb,b}$.  
This yields
\be
\frac{\Delta p_b}{p_b} = 
	\left (\frac{m_a}{m_b} \right)
	\left [\frac{\Delta v(r)}{v_b(r)} \right]
	\left (\frac{f_b P_{\rm orb,b}}{t_c} \right) 
	\lesssim \frac{P_{\rm orb,b}}{t_b}.
\ee
The second condition requires that the collisional influx into the LC
must be at a high-enough rate that the magnetosphere receives a continuous
flow of evaporated material during a burst.   
Assuming collision fragments are directed
isotropically, the fraction that enters the LC is
$f_{\rm LC} \approx \left( cP/4\pi r \right)^2$, yielding a mean time
between injections of $t_c/f_{\rm LC}$ that we require to be
less than the evaporation time of an asteroid, $t_{\rm evap}$. 
Taken together and using $\Delta v(r) / v_b(r) \sim e/2$, we can bracket
the collision time using
\be
\frac{e f_b t_b}{2} \left(\frac{m_a}{m_b} \right) \le t_c 
	\le f_{\rm LC} t_{\rm evap}.
\ee 

Evaluating for $m_a/m_b = (R_a / R_b)^3 \approx 10^{-15}$ (e.g. for
a 100 km cue-ball and $R_a = 1$~m),  assuming that collisions
take place at periastron, $r = a(1-e)$, and using an evaporation time
$t_{\rm evap} = 10^4 t_{\rm evap, 4}$~s 
(which applies to meter-size rocks for a variety of temperatures), we obtain
\be
10^{-4.1}\, {\rm s}\, f_{\rm b, 5}\left ( \frac{m_b/m_a}{10^{15}} \right)
\le t_c 
\le 10^{-2.4}\, {\rm s}\, t_{\rm evap, 4}.
\ee
This demonstrates that the orbital configuation can be maintained for
a sufficiently long time that we do not require ourselves to be 
observing at a special time.   


\subsection{Other possibilities}

Analogs to other solar-system phenomena may apply, though they do
not readily yield a quasi-periodicity in the magnetospheric response.
For example, ``horseshoe'' orbits of small asteroids that enclose
the $L_3- L_5$ Lagrangian points in a NS-large-asteroid system
have libration periods that can be large multiples of the orbital
period.   These orbits span a range of distances  from the NS that
is wider at the $L_{4,5}$ points than at the $L_3$ point.   For orbits
that are closer to the NS, one might expect an increase in collisions
or evaporation rate that could enhance injection into the LC.
Only if the orbits are populated with a small number of objects would
we expect to see episodic or quasi-periodic effects, however. 
    
As another variant, the 40~d quasi-period may be associated with 
the period of interaction
between two orbiting objects in resonance.    
Mean-motion resonances,
such as those that produce gaps within the solar system's asteroid belt
and in planetary ring systems, induce chaotic motions that could drive
asteriods into the NS.  However, this process is likely to produce
a more continuous and possibly large collision rate and, hence, 
less-intermittent bursts. Consequently, we favor the cue-ball model
for producing the quasi-periodic effects in B1930+24. 

\subsection{Rotating Radio Transients (RRATs)}\label{sec:rrats}

RRAT objects identified to date populate the same part of the $P-\Pdot$
diagram as objects that show prominent nulling, with the possible exception
of J$1819-1458$, which has a large spindown-derived magnetic field that
is similar to that of magnetars.   We consider RRATs to be associated
with NS that are similar or identical to canonical pulsars.  They are
observed to be quiescent most of the time either because they are
intrinsically so or because their radio beams do not typically point
toward us.    RRAT pulses have duty cycles that are significantly smaller
than those of canonical pulsars \cite{m+06} with comparable periods,
suggesting that the intermittent emission is from a region of the 
magnetosphere smaller than typical.

If RRAT objects are active pulsars not usually beamed toward us, their intermittency could be explained by the injection asteroidal material into the outer magnetosphere. 
The pair production rate would then be enhanced, which would increase the overall plasma density and the altitude
of emission.  As discussed earlier, this would increase the probability
of detection and make some objects intermittently detectable.  
Pulses from such
objects would be {\it wider} than typical if the entire open field line
region is affected.  Only if a restricted region were affected might
we expect to see narrow pulses.  This appears to be the case
with canonical pulsars for which single pulses are almost always
narrower than the average of many pulses, signifying that the
open field line region is not instantaneously filled with
coherent emitters. 

A perhaps more reasonable alternative is that asteroidal 
material triggers an inactive outer gap.  It is easier to inject 
material into an
outer gap and such gaps should exist in all pulsars.  In long period
pulsars, however, outer gaps have potential drops too small 
to sustain self-generated pair cascades \cite[][]{zcjl04}.  However, 
external triggers can drive pair avalanches in such gaps.  
It is noteworthy that
for objects where outer-gap emission is a plausible candidate
source of radio emission, the radio pulse components are quite narrow.
The best cases are the Crab pulsar and the MSP B1937+21, both of
which show high-energy emission that aligns in pulse phase (or nearly
so) with the radio components.  These two objects show giant pulses
in these same components.   Giant pulses from the Crab pulsar are
very narrow if one corrects for pulse broadening from radio-wave
scattering, e.g. $W/P \sim 100\,\mu s\, / 33\,{\rm ms} \sim 0.3$\%. 
The pulses from B1937+21 \cite{jap01} are a wider $\sim 2.5$\%.
By triggering the outer gap,  pair production and coherent emission
may be produced directly and we may see the emission from either
the ingoing or outgoing beam.  In this regard, our model is
similar to that of \cite{zgd06}, who postulate intermittency
and reversal of beaming direction but did not propose a specific
mechanism for the intermittency, as we do here.   Alternatively
(or in addition), 
activation of the outer gap may induce particle flows that
in turn activate or alter processes lower in the magnetosphere,
such as the polar cap.  

Our model implies that a triggered avalance would continue
for only a finite amount of time.  However, it can be longer than
a NS spin period if accretion is sustained. 

In some of the RRATS we may expect to see polar cap emission from the pulsar.
Thus, a test of the model would be to search for this emission.  We would expect to see these components at quite different pulse phases. 

  
\section{Tests and Further Applications }\label{sec:tests}

We have suggested a number of plausbility arguments in favor of
our model, including 
the distribution of nulling and RRAT pulsars
in the $P-\Pdot$ diagram (Figure~\ref{fig:ppdot}), 
the temperature and spin rate regime in
which we expect disk material can enter a pulsar's magnetosphere
in a  largely neutral state (Figure~\ref{fig:Rvsr}), and 
the distribution of spin-magnetic-moment
angles in Figure~\ref{fig:null_angle}.  
Here we consider ways in which the model can be tested and used
to infer the properties of material orbiting radio pulsars.

As possible tests, we suggest that identification of counter-examples
to the trends seen in the $P-\Pdot$ diagram would be valuable.  For example,
identifying nulls in MSPs or in young pulsars with high thermal luminosity
or high magnetospheric luminosity combined with favorable orientation
would suggest alternative causes in those objects.   In the case of
B1931+24, the pulsar showing quasi-periodic bursts with correlated
increase in $\Pdot$, we suggest that there may be multiple orbiting
objects that may be detectable as arrival time perturbations that must
be disentangled from the changes in $\Pdot$.   A careful investigation
of the ``timing noise'' in a large sample of pulsars, such as that
presented in \cite{h+04b}, may reveal trends in types of 
timing noise vs. $P$ and $\Pdot$ and may corroborate the existence
of torque fluctuations in those objects that also show nulls and other
evidence for accretion.

\subsection{Young Pulsars}

Injection of charged particles through asteroid intrusion is much less
likely from young pulsars that have small magnetospheres (if they have
low spin periods), have hot surface temperatures ($\gtrsim 10^6$~K),
and have high non-thermal luminosities.   Exceptions may exist
if occasional, large asteroids on low-angular-momentum trajectories
can rapidly get inside the LC before they are evaporated.    Thus,
the division between ``young and hot'' and ``old and cool'' NS may 
depend on the collision rate and size of the asteroid disk, which will
determine the rate at which events might be seen from a given object.   
Already, the trend appears to be that objects near the death line
in the $P-\Pdot$ diagram are much more prone to showing transient effects
that can be attributed to asteroids.  Nonetheless, we may expect
to see occasional departures from this trend.    A notable case
may be RRAT J$1819-1458$, which has a small spindown age (117 kyr)
but has a long period (4.3~s) \cite[][]{m+06}.  Thus its LC is large but its 
surface temperature \cite[][]{r+06}
 is also large, $T_* \approx 0.12$~keV~$\approx 10^{6.1}$~K.   
Larger asteroids ($\gtrsim 100$~m) can enter, though not deeply,
into the magnetosphere before they evaporate, allowing the possiblity
that they trigger a dormant outer-gap region.   

\subsection{Millisecond Pulsars}

Most MSPs  have white-dwarf companions in compact
orbits, so those objects are unlikely to have additional debris
given their large ages ($0.1 - 10$ Gyr).  
Disks may exist around isolated MSPs as condensed evaporant
from the companion star that once provided accreting material 
responsible for spinup of the MSP.   Such disks may disappear on time
scales much shorter than MSP radio-emitting lifetimes.  
For cases where there are disks,   
MSPs are unlikely to allow injection of asteroids into their
magnetospheres  because their light cylinders are very small,
$r_{\rm LC} = 10^{7.4}\,{\rm cm}\,(P/5\,{\rm ms})$, and the non-thermal
radiation flux typically is high. 

\subsection{Infrared Disk Detection}

Unlike the recent detection of a fairly massive disk around the 
magnetar 4U 0142+61 \cite[][]{wck06},  
disks around canonical radio pulsars  will be much more difficult to
detect.  

\cite{bryden06} recently observed pulsar planetary system B1257+12 in the 
mid-infrared looking for asteroidal dust.  No evidence was found for 
emission, but the limits do not rule out a disk comparable to the 
Solar System's asteroid belt.  We expect the planetary system around B1257+12 
is less likely to contain a massive asteroid belt, 
because gravitational interactions will remove debris at a much faster rate
than for an asteroid belt without planets.
For the objects we consider, 
the disk is much more compact about the central source, so the 
emission should peak in the near infrared.  

We assume that the disk, with mass $M_a$, is composed of 
spherical asteroids of area $r_a$, density $\rho$, and albedo A.  
The disk is heated by the thermal radiation of the neutron star 
(radius $R_{*}$, surface temperature $T_{*}$).  We also assume the disk is orthogonal to our line of sight.  
The temperature $T_a$ of the asteroid at distance r from 
the NS is given by Eq.~\ref{eq:Ta}.
 As the disk is very small,  an observer 
will see the disk as a point source.  
To calculate the total flux density, we add the flux directed towards 
us from all the asteroids:
\begin{equation}
F_{\nu} = 
\int_{r_1}^{r_2}\, dr 
\pi r_a^2 
F_{\nu,{\rm em}}(T_a)
\frac{dN}{dr}, 
\end{equation}  
where $dN/dr = 2\pi r n_a \epsilon_h r_2$ is the number vs. $r$ and
we assume uniform number density, $n_a$. 
Adding up the Planck flux from each asteroid, we obtain
\begin{eqnarray}
F_{\nu} = 
\frac{45}{8 \pi^5 d^2} 
\frac{\sigma_{\rm SB} (1-A)T_{*}^4}{\nu} 
\frac{R_{*}^2}{r_a r_2^2} \frac{M_a}{\rho} 
\int_{x_1}^{x_2}
\frac{dx\,x^3}{e^x -1},
\end{eqnarray}
where
$x_{1,2} = (h \nu/ k T_*) \sqrt{2 r_{1,2}/R_{*}}$.
Using our standard disk parameters  (Eq.~\ref{eq:Sigma}-\ref{eq:Na}),
the  peak flux occurs at a wavelength 
$\lambda \approx 13 \mu{\rm m} T_5^{-1} R_6^{\frac{1}{2}} (r_{2,-2} r_{1,10})^{\frac{1}{4}}$, 
when the peak of the integrand 
($x \approx 2.8$) occurs at the geometric average of the two endpoints.  The 
integral is approximately 3 for  
the standard parameters but could be as high as  
$\approx 6$ for asymptotically large disks, and will be much smaller 
for narrower disks.  However, narrower disks would not satisfy the 
contraints suggested in \S \ref{sec:obscons}.  
We find then that the observed flux density is
\be
F_{\nu} = 10^{-1.4} {\rm \mu Jy} \, 
d_{\rm kpc}^{-2} T_{5}^3 R_6^{\frac{3}{2}} 
M_{\rm a,4} r_{\rm a,2}^{-1} r_{2,-2}^{-\frac{7}{4}}r_{1,10}^{\frac{1}{4}} 
\rho^{-1}.
\ee
Though this fiducial value would require long exposures on today's best 
instruments\footnote{The Spitzer IRAC point-source sensitivity
is 0.6\,$\mu$Jy
(1$\sigma$) for a 100 s exposure at 3.6 $\mu$m 
(\url{http://ssc.spitzer.caltech.edu/irac/})},  
the uncertainty in the disk parameters allows a significant increase 
in fluence.  Furthermore, the stong sensitivity to the temperature of 
the neutron star could allow for disks to be detected surrounding 
younger, hotter NSs.   

\subsection{Radio Disk Detection}

Under suitable conditions, an asteroid belt may be detected through
reflections of pulsed radio emission from asteroids larger than
the wavelength \cite[][]{p93}.   
The ratio of reflected to pulsed
flux is approximately 
\be
\frac{S_u}{S_p} &=& \frac{3 M_a A}{4\pi r^2 \rho} f_g \tilde R_{\rm a}^{-1} \\
	\tilde R_{a} &=& \frac {\langle R_a^3 \rangle} {\langle R_a^2 \rangle}, 
\ee   
where $f_g$ is a geometrical factor that takes into account beaming
of the pulsed flux and the solid angle into which pulsed flux is reflected; 
$\langle R_a^x \rangle$ is the average over a distribution
of asteriod sizes for sizes larger than
the wavelength, implying that the mass $M_a$ in the belt also includes
only those objects.  In his analysis, \cite{p93} assumed a fan beam
for the pulsar radiation with $f_g = 1$;  for many pulsars, a pencil beam
appears to be more appropriate, leading to $f_g < 1$.   
For nominal parameters we have
\be
\frac{S_u}{S_p} =  10^{-5.2}\, 
	\frac{f_g A_{0.1} M_{\rm a, -4}} {\rho r_{10}^2 \tilde R_{\rm a, km}}.
\ee  
For strong pulsars with $S_p\approx 1$~Jy, the unpulsed reflected flux
$\sim 10\,\mu$Jy.  With present instrumentation this is very
difficult to measure while with the EVLA and the
Square Kilometer Array, it will be
straightforward to detect  or to place highly constraining limits. 

\subsection{High-energy Studies}


Canonical radio pulsars are easily detected in X-rays through their
magnetospheric emission and, in some cases, their thermal surface
radiation.   For cases where an acceleration region is reactivated by
an asteroid, as is plausibly the case for the RRAT objects, we may expect
to also see high-energy radiation.         

\subsection{Interstellar Comets}

While no interstellar comet (ISC) has yet been identified, the Oort cloud's
existence implies that a large number of ISCs should exist
\cite[e.g.][]{s90}.  A typical
density may be $n_c = 10^{14}n_{14}\,{\rm pc^{-3}}$, so for a
characteristic velocity of 30 km s$^{-1}$ and a cross section for
an asteroid disk of $\pi r_{2}^2 = 10^{-4} \pi r_{2,-2}^2\,{\rm AU^2}$,
the typical collision time is
\be
t_c = 10^{4.6}\,{\rm yr}\, \left (n_{14}r_{2,-2}^2 v_{30} \right)^{-1},
\ee 
which is too long to be of interest for our standard disk. However,
a dense disk extending to $r_2 \sim 1$~AU and having
high optical depth would show an event about once per year.     
Comet-disk collisions would be brief but might induce
highly episodic behavior in some objects.

\subsection{Pulsar Dependent Disk Properties}

The accretion model can be used as the basis for inferring the presence
and properties of  disks from the emission properties of various objects.
Time-variable effects appear in some objects and not others, suggesting
that there is a wide range of disk masses  among pulsars and a wide
range in instrinsic pulsar properties that regulate the inflow 
of neutral material.    A variation in disk properties is not surprising
because pulsars comprise a runaway stellar population whose velocities
originate from momentum kicks imparted at or shortly after the formation
of the NS \cite[e.g.][]{lcc01}.   The amount of fallback material that
ultimately settles into a disk will depend on the amplitude of the
natal kick.  A correlation study of NS space velocities and their
radio intermittency may provide additional confirmation of the model
and constraints on disk properties.






\section{Summary and Conclusions}\label{sec:summary}

We have presented a new model that highlights the effects that 
asteroids in circumpulsar disks will have on electrodynamic and
radiation processes in the magnetospheres of neutron stars.   
Implied phenomena may occur in both active radio pulsars and those that
are usually quiescent.   
The key element is
the injection of asteriods as macroscopic objects into
pulsar magnetospheres, whereupon they evaporate and provide charges
that are further processed.  The rich set of consequences of this
picture unifies what is otherwise a dizzying array of phenomena
--- drifting subpulses, pulse-shape mode changes, nulling, and bursts. 
Support for the model includes the theoretical expectation that orbital
debris will accumulate in supernova fallback disks and that 
large asteroids can survive for much longer than the typical
radio lifetime of a radio pulsar.   Collisions and radiation-driven
migration --- processes identified in the solar system --- inject asteriods 
into the light cylinder and provide charges at rates that 
perturb acceleration regions and presumably radio-emission.   Objects
that show bursts or termination of emission (nulls) with short
rise and fall times appear to have spin periods and period
derivatives that are consistent with objects that are expected to 
be cool and have large magnetospheres, just those for which asteroidal
contamination is the most likely.   Empirical confirmation
of the model would involve detection of debris disks around pulsars
that show intermittency and mass-constraining non-detections for those
that do not.  Unfortunately, disk detection is unlikely with
existing instrumentation and may require the EVLA and the SKA to detect
reflected pulsar radiation in the radio band or a large infrared instrument
for direct detection.   The quasi-periodic burster, B1931+24, may provide
the best opportunity with existing telescopes  because the elliptical
orbit in our model suggests the presence of additional objects that
may produce orbital recoil of the pulsar measureable in pulse timing
data.

Further studies are warranted to explore asteroidal induced 
torque variations and timing noise, inference of disk properties from  
radio-emission properties, multiwavelength studies of objects
that do and do-not show associated radio phenomena, and theoretical
study of fallback disks as a function of supernova kick amplitudes. 
Correlation studies will test the model and point to
other information-gleaning methods, such as exploring the relationship
between the spin/magnetic-moment angle and the occurrence of
nulls  and finding exceptional objects,
such as young, hot neutron stars that show nulls and appear to 
contradict the model.  The spin history may determine which objects
retain asteroid belts, as those NS born with high magnetic field strengths
and high spin rates will significantly ablate asteroids.  However, 
a dense disk may rapidly spin down the NS and allow asteroids to form.     
The durations of nulls and bursts are 
closely related to the lifetimes of asteroids as they enter a
pulsar magnetosphere.  The lifetime, in turn, depends on radiative
and particle conditions within the magnetosphere, so observable
effects may provide robust contraints on NS temperatures, radiation
mechanisms,  particle flows, and NS masses via cooling mechanisms
that are mass dependent.    

\setcounter{footnote}{0}

\acknowledgments

We thank Joe Burns, Michael Kramer, Richard Lovelace, Andrew Lyne, 
Maura McLaughlin, Marina Romanova, and Ira  Wasserman 
for helpful conversations.
This work was supported by the National Science Foundation
through grant AST-02-06035 to Cornell University.

\appendix

\section{Induction Heating}

We consider two types of induction heating, one from the temporally changing
magnetic field experienced by an asteroid that has no net current flow
through it, the other from a field-induced current through the asteroid.

Well outside the corotation radius, 
$r_{co} = [GM_{\rm *} P^2/ (2\pi)^2 ]^{1/3}\approx10^{8.2}\,{\rm cm}\,P^{2/3}$,
the asteroid is subjected to a changing magnetic field,
$\partial B/\partial t$, that is determined
by the NS spin; this tends to zero as the asteroid approaches
corotation but most asteroids will evaporate before reaching corotation.
The generated electric field $\propto \partial B /\partial t$ in the
asteroid's frame is dissipated within a skin depth $\delta \ll R_a$, 
yielding an equilibrium temperature
(with $\sigma$ the conductivity and 
$\sigma_{\rm SB}$ the Stefan-Boltzmann constant)
\be
T_{a, \sigma} = \left (\frac{c}{24\pi \sigma_{\rm SB}} \right )^{1/4}
	\left (P\sigma \right)^{-1/8} B^{1/2}
\approx  10\,K\, 
	P^{-1/8}
	\left ( B_{12}\sin\alpha \right)^{1/2} r_{10}^{-3/2}  
	\left(\frac{\sigma}{\sigma_{\rm max}}\right)^{-1/8}, 
	\label{eq:induction}
\ee   
where $\sigma/\sigma_{\rm max}$ is the conductivity relative
to a maximum that is taken to be that 
of pure copper, $\sigma_{\rm Cu}\approx 5\times 10^{17}$~s$^{-1}$.  
Asteroidal material will have lower conductivities and thus higher temperatures
than implied in the equation.  However,  the skin depth
$\delta = (c / 2\pi) \sqrt{P/\mu \sigma}$ ($\sim 7\sqrt P$ cm for copper) 
also increases with decreasing conductivity so that an asteroid's 
resistance reaches a maximum when $\delta \sim R_{\rm a}$.  This limits
the heating and thus the temperature coefficient 
to only a factor of a few larger than in    
Eq.~\ref{eq:induction}. 
Ohmic heating increases faster with
decreasing radius than radiative heating and would be competitive
with radiative heating at radii $\sim 10^{8}$ cm if asteroids could reach
that radius (and ignoring corotation).

The electric field in the frame of an asteroid moving 
orbitally or ballistically near the light cylinder of a pulsar 
with $v\ll c$, is 
$\mathbf{ E} = (\mathbf{\Omega \times r) \times {B}}/ c$, 
or approximately 
$10^{5.4}\,{\rm volts\,\,m^{-1}} B_{12}(r_{\rm LC}/r)^2P^{-3}R_{*,6}^3$.  
If a circuit is established through the magnetosphere and 
asteroid, the current density
$\mathbf {J} = \sigma \mathbf{E}$ yields ohmic heating power 
$\tilde P \approx {R_a^3 JE}$ as long as the
induced azimuthal magnetic field at the asteroid surface satisfies
$B_i \approx  2\pi R_a J/c \lesssim B$.  The implied maximum
current density,   $J_{\rm max} \approx c  B / 2\pi R_a$,  
corresponds to an evaporation flux,
\be
\frac{d\dot N}{dA} = \frac{J_{\rm max}}{e} 
	\approx 
	10^{18.0} \,{\rm s^{-1}\,cm^{-2}}
	B_{12} P^{-3} R_{*,6}^3 R_{a,2}^{-1} (r_{\rm LC} / r)^3.
\ee
and a circuit resistance 
$\sim 4r/cr_{\rm LC}$ (cgs) or about  $120(r/r_{\rm LC})$ Ohms. 
The maximum heating power is
\be
\tilde P_{\rm max} \approx
	c (R_a B)^2 (r/\rlc) 
\approx 
10^{16.4}\,{\rm erg\,\,s^{-1}}\,
\left  [P^{-3} B_{12} R_{a,2} R_{*,6}^3 \right ]^2 (r_{\rm LC} / r)^5. 
\ee
When the current density is less than $J_{\rm max}$,
the heating rate is $\tilde P = (J/J_{\rm max})\tilde P_{\rm max}$.
The corresponding equilibrium temperature is 
\be
T_{a,i} \approx \left(
		\frac{\tilde P} 
			{4\pi \sigma_{\rm SB} R_a^2}
		\right )^{1/4} 
	\approx 10^{3.9}\,{\rm K}\, R_{a,2}^{-1/2} 
		\left( 
		  \frac{\tilde P_{\rm max}}{10^{16.4} \,{\rm erg\,s^{-1}}}
	 	\right)^{1/4}
		\left(
		  \frac{\tilde P}{\tilde P_{\rm max}}
		\right)^{1/4}.
\ee 
As an asteroid first enters the magnetosphere at 
$r\sim r_{\rm LC}$ of a long-period pulsar, it is too cool to 
provide the necessary particle flux
(c.f. Figure~\ref{fig:evap12}),
so heating is dominated by radiative effects (see text) until evaporation
becomes large enough to supply the current.    For NS
with $10^5$~K surface temperatures and low magnetic fields, injected asteroids 
reach $\sim 10^9$~cm
before they have 2000 K temperatures 
(Eq.~\ref{eq:Ta} and Figure~\ref{fig:Rvsr}), at which their
evaporation rates  are too low to provide the full current density.
However, with a canonical field strength of $10^{12}$~G,
the asteroid temperature and evaporation rise very rapidly with 
decreasing radius so that runaway heating     
drives an evaporation avalanche, exploding the asteroid.
This occurs at about $10^{9.3}$~cm for a meter-sized asteroid.
For pulsars with smaller field strengths, this radius will decrease.
These numbers are altered if there is significant particle flux
from elsewhere in the magnetosphere that provides a significant amount
of current through the asteroid.       

\clearpage
\LongTables
\begin{deluxetable}{rrrrrrrr} 
\tabletypesize{\footnotesize}
\tablecolumns{8} 
\tablewidth{0pc} 
\tablecaption{Parameters for Nulling and Nulling-Limited Pulsars 
	\label{tab:nulldata}
} 
\tablehead{ 
 \colhead{Name (J2000)} & \colhead{Name (B1950)}   & \colhead{$P$}    & \colhead{$\log \dot{P}$} & 
\colhead{$\log \tau_s$}    & \colhead{Nulling Fraction}  & 
	\colhead{$\alpha$} & \colhead{References} \\
 & & \colhead{(s)} & \colhead{(s s$^{-1}$}) & \colhead{(yr)} & \colhead{(\%)} &
	\colhead{(deg)}  
}

\startdata 
\\
J1744$-$3922 & \nodata & 0.1724 &  $-$17.8 & 9.25 & 75 $\pm$ 32 & \nodata & F04, \nodata \\        
J0828$-$3417 & B0826$-$34 & 1.8489 &  $-$15.0 & 7.48 & 70 $\pm$ 35 & 5.0 & B92, Upper   \\    
J1944+1755 & B1942+17 & 1.9969 &  $-$15.1 & 7.64 & 60 & \nodata & B92, \nodata \\        
J1115+5030 & B1112+50 & 1.6564 &  $-$14.6 & 7.02 & 60 $\pm$ 5 & 32 & B92, R93   \\    
J1819+1305 & \nodata & 1.0604 &  $-$15.4 & 7.67 & 50  & \nodata & LCX02, \nodata \\        
\\
J1946+1805 & B1944+17 & 0.4406 &  $-$16.6 & 8.46 & 50 $\pm$ 7 & \nodata & V95, R93 \\        
J0034$-$0721 & B0031$-$07 & 0.9430 &  $-$15.4 & 7.56 & 44.6 $\pm$ 1.3 & 6 & V95, R93   \\    
J0754+3231 & B0751+32 & 1.4423 &  $-$15.0 & 7.33 & 34 $\pm$ 0.5 & 26 & B92, R93   \\    
J1649+2533 & \nodata & 1.0153 &  $-$15.3 & 7.46 & 30  & \nodata & L04, \nodata \\        
J0528+2200 & B0525+21 & 3.7455 &  $-$13.4 & 6.17 & 25 $\pm$ 5 & 21 & B92, MR02   \\    
\\
J2321+6024 & B2319+60 & 2.2565 &  $-$14.2 & 6.71 & 25 $\pm$ 5 & 18 & B92, R93   \\    
J1945$-$0040 & B1942$-$00 & 1.0456 &  $-$15.3 & 7.49 & 21 $\pm$ 1 & 26 & B92, R93   \\    
J1136+1551 & B1133+16 & 1.1880 &  $-$14.4 & 6.70 & 15.0 $\pm$ 2.5 & 46 & B92, MR02   \\    
J2113+4644 & B2111+46 & 1.0147 &  $-$15.1 & 7.35 & 12.5 $\pm$ 2.5 & 9 & B92, R93   \\    
J0942$-$5552 & B0940$-$55 & 0.6644 & $-$13.6 & 5.66 & $\leq$ 12.5 & 25 & B92, R93  \\                        
\\
J2330$-$2005 & B2327$-$20 & 1.6436 &  $-$14.3 & 6.75 & 12 $\pm$ 1 & 60 & B92, R90   \\    
J0659+1414 & B0656+14 & 0.3849 &  $-$13.3 & 5.05 & 12 $\pm$ 4 & 30.0 & B92, R93   \\    
J1057$-$5226 & B1055$-$52 & 0.1971 & $-$14.2 & 5.73 & $\leq$ 11 & 90.0 & B92, B93  \\                        
J0304+1932 & B0301+19 & 1.3876 &  $-$14.9 & 7.23 & 10 $\pm$ 5 & 38 & B92, MR02   \\    
J2305+3100 & B2303+30 & 1.5759 &  $-$14.5 & 6.93 & 10 $\pm$ 3 & 20.5 & RWR05, R93;   \\    
\\
J1900$-$2600 & B1857$-$26 & 0.6122 &  $-$15.7 & 7.68 & 10 $\pm$ 2.5 & 25 & B92, R93   \\    
J2048$-$1616 & B2045$-$16 & 1.9616 &  $-$14.0 & 6.45 & 10.0 $\pm$ 2.5 & 34 & V95, MR02   \\    
J0943+1631 & B0940+16 & 1.0874 &  $-$16.0 & 8.28 & 8 $\pm$ 3 & 53 & B92, R93   \\    
J2157+4017 & B2154+40 & 1.5253 &  $-$14.5 & 6.85 & 7.5 $\pm$ 2.5 & 20 & B92, R93   \\    
J0837+0610 & B0834+06 & 1.2738 &  $-$14.2 & 6.47 & 7.1 $\pm$ 0.1 & 30 & B92, MR02   \\    
\\
J0944$-$1354 & B0942$-$13 & 0.5702 & $-$16.3 & 8.30 & $\leq$ 7 & 45 & V95, R93  \\                        
J1532+2745 & B1530+27 & 1.1248 &  $-$15.1 & 7.36 & 6 $\pm$ 2 & 26 & B92, R93   \\    
J1239+2453 & B1237+25 & 1.3824 &  $-$15.0 & 7.36 & 6.0 $\pm$ 2.5 & 53 & B92, MR02   \\    
J0151$-$0635 & B0148$-$06 & 1.4647 & $-$15.4 & 7.72 & $\leq$ 5 & 14.5 & B92, R93  \\                        
J2022+5154 & B2021+51 & 0.5292 & $-$14.5 & 6.44 & $\leq$ 5 & 23 & B92, R93  \\                        
\\
J1926+0431 & B1923+04 & 1.0741 & $-$14.6 & 6.84 & $\leq$ 5 & 34.5 & B92, R93  \\                        
J1614+0737 & B1612+07 & 1.2068 & $-$14.6 & 6.91 & $\leq$ 5 & 24.5 & B92, R93  \\                        
J0953+0755 & B0950+08 & 0.2531 & $-$15.6 & 7.24 & $\leq$ 5 & 74.6 & V95, EW01  \\                        
J1841+0912 & B1839+09 & 0.3813 & $-$15.0 & 6.74 & $\leq$ 5 & 86.1 & B92, EW01  \\                        
J0826+2637 & B0823+26 & 0.5307 & $-$14.8 & 6.69 & $\leq$ 5 & 81.1 & B92, EW01  \\                        
\\
J1910+0358 & B1907+03 & 2.3303 &  $-$14.3 & 6.92 & 4.0 $\pm$ 0.2 & 6.0 & B92, R90   \\    
J1243$-$6423 & B1240$-$64 & 0.3885 & $-$14.3 & 6.14 & $\leq$ 4 & 33 & B92, R93  \\                        
J1456$-$6843 & B1451$-$68 & 0.2634 & $-$16.0 & 7.63 & $\leq$ 3.3 & 37.0 & B92, R93  \\                        
J2317+2149 & B2315+21 & 1.4447 &  $-$15.0 & 7.34 & 3.0 $\pm$ 0.5 & 88 & B92, R93;   \\    
J2022+2854 & B2020+28 & 0.3434 & $-$14.7 & 6.46 & $\leq$ 3 & 56 & B92, MR02  \\                        
\\
J0152$-$1637 & B0149$-$16 & 0.8327 & $-$14.9 & 7.01 & $\leq$ 2.5 & 84 & V95, R93  \\                        
J2219+4754 & B2217+47 & 0.5385 & $-$14.6 & 6.49 & $\leq$ 2 & 42 & B92, R93  \\                        
J0814+7429 & B0809+74 & 1.2922 &  $-$15.8 & 8.09 & 1.42 $\pm$ 0.02 & 9 & B92, R93   \\    
J0837$-$4135 & B0835$-$41 & 0.7516 & $-$14.5 & 6.53 & $\leq$ 1.2 & 50 & B92, R93  \\                        
J0820$-$1350 & B0818$-$13 & 1.2381 &  $-$14.7 & 6.97 & 1.01 $\pm$ 0.01 & 15.5 & B92, R93   \\    
\\
J1932+1059 & B1929+10 & 0.2265 & $-$14.9 & 6.49 & $\leq$ 1 & 35.97 & B92, EW01  \\                        
J2116+1414 & B2113+14 & 0.4402 & $-$15.5 & 7.38 & $\leq$ 1 & 30.0 & B92, R93  \\                        
J1820$-$0427 & B1818$-$04 & 0.5981 & $-$14.2 & 6.18 & $\leq$ 0.75 & 65 & B92, R93  \\                        
J1948+3540 & B1946+35 & 0.7173 & $-$14.2 & 6.21 & $\leq$ 0.75 & 32.0 & B92, R90  \\                        
J1752$-$2806 & B1749$-$28 & 0.5626 & $-$14.1 & 6.04 & $\leq$ 0.75 & 42 & B92, R93  \\                        
\\
J2055+3630 & B2053+36 & 0.2215 & $-$15.4 & 6.98 & $\leq$ 0.7 & 34 & B92, R93  \\                        
J0452$-$1759 & B0450$-$18 & 0.5489 & $-$14.2 & 6.18 & $\leq$ 0.5 & 24.0 & B92, R93  \\                        
J1913$-$0440 & B1911$-$04 & 0.8259 & $-$14.4 & 6.51 & $\leq$ 0.5 & 64.0 & B92, R93  \\                        
J1823+0550 & B1821+05 & 0.7529 & $-$15.6 & 7.72 & $\leq$ 0.4 & 32 & B92, R93  \\                        
J0738$-$4042 & B0736$-$40 & 0.3749 & $-$14.8 & 6.57 & $\leq$ 0.4 & 17.0 & B92, R93  \\                        
\\
J1644$-$4559 & B1641$-$45 & 0.4551 & $-$13.7 & 5.56 & $\leq$ 0.4 & 33 & B92, R93  \\                        
J0630$-$2834 & B0628$-$28 & 1.2444 & $-$14.1 & 6.44 & $\leq$ 0.3 & 13.5 & B92, R93  \\                        
J2018+2839 & B2016+28 & 0.5580 & $-$15.8 & 7.78 & $\leq$ 0.25 & 39 & B92, R93  \\                        
J1921+2153 & B1919+21 & 1.3373 & $-$14.9 & 7.20 & $\leq$ 0.25 & 34 & B92, MR02  \\                        
J1534$-$5334 & B1530$-$53 & 1.3689 &  $-$14.9 & 7.18 & $\leq$ 0.25 & \nodata &  B92, PCE\\                        
\\
J0332+5434 & B0329+54 & 0.7145 & $-$14.7 & 6.74 & $\leq$ 0.25 & 32 & B92, MR02  \\                        
J1645$-$0317 & B1642$-$03 & 0.3877 & $-$14.7 & 6.54 & $\leq$ 0.25 & 70 & B92, R93  \\                        
J0742$-$2822 & B0740$-$28 & 0.1668 & $-$13.8 & 5.20 & $\leq$ 0.2 & 37 & B92, R93  \\                        
J1844+1454 & B1842+14 & 0.3755 & $-$14.7 & 6.50 & $\leq$ 0.15 & 29 & B92, R90  \\                        
J1919+0021 & B1917+00 & 1.2723 & $-$14.1 & 6.42 & $\leq$ 0.1 & 81 & B92, R93  \\                        
\\
J1745$-$3040 & B1742$-$30 & 0.3674 & $-$14.0 & 5.74 & $\leq$ 0.1 & 51 & B92, R93  \\                        
J1731$-$4744 & B1727$-$47 & 0.8298 & $-$12.8 & 4.91 & $\leq$ 0.1 & 64 & B92, R93  \\                        
J1607$-$0032 & B1604$-$00 & 0.4218 & $-$15.5 & 7.34 & $\leq$ 0.1 & 50 & B92, MR02  \\                        
J0823+0159 & B0820+02 & 0.8649 & $-$16.0 & 8.12 & $\leq$ 0.06 & 46 & B92, R93  \\                        
J0525+1115 & B0523+11 & 0.3544 & $-$16.1 & 7.88 & $\leq$ 0.06 & 78 & B92, R93  \\                        
\\
J1935+1616 & B1933+16 & 0.3587 & $-$14.2 & 5.98 & $\leq$ 0.06 & 72.0 & B92, R93  \\                        
J1430$-$6623 & B1426$-$66 & 0.7854 & $-$14.6& 6.65 & $\leq$ 0.05 & \nodata &  B92, PCE\\                        
J0922+0638 & B0919+06 & 0.4306 & $-$13.9 & 5.70 & $\leq$ 0.05 & 48.0 & B92, R93  \\                        
J2046+1540 & B2044+15 & 1.1383 & $-$15.7 & 8.00 & $\leq$ 0.04 & 40 & B92, R93  \\                        
J1559$-$4438 & B1556$-$44 & 0.2571 & $-$15.0 & 6.60 & $\leq$ 0.04 & 32 & B92, R93  \\                        
\\
J1740+1311 & B1737+13 & 0.8031 & $-$14.8 & 6.94 & $\leq$ 0.02 & 41 & B92, R93  \\                        
J0629+2415 & B0626+24 & 0.4766 & $-$14.7 & 6.58 & $\leq$ 0.02 & 30 & B92, R93  \\                        
J0437$-$4715 & \nodata & 0.0058 &  $-$19.2 & 9.20 & $\leq$ 0.0016 & \nodata &  B92, \nodata\\                        
J0835$-$4510 & B0833$-$45 & 0.089 & $-$12.9 & 4.05 & $\leq$ 0.0008 & 90 & B92, R93  \\                        
\enddata
\tablecomments{A list of pulsars known to null and pulsars for which limits have been placed on nulling fraction.  The period $P$, period derviative $\dot{P}$, the spin-down age $\tau_s$, and the angle between the spin and magnetic axes $\alpha$ are also given.  The first reference is for the nulling fraction, and second is $\alpha$, the angle between the magnetic and spin axes.   R90 refers to \cite{rankin1990}, B92 to \cite{b92},  R93 to \cite{rankin1993}, V95 to \cite{1995MNRAS.274..785V},  LCX02 to \cite{lor02}, MR02 to \cite{mr02}, EW01 to \cite{ew01},  L04 to \cite{l+04}, F04 to \cite{fsk+04}, and RWR05 to \cite{2005MNRAS.357..859R}}.
\end{deluxetable}
\end{document}